\documentclass[apj]{emulateapj-rtx4}
\usepackage{graphicx}
\usepackage{times,epsfig,natbib,amssymb,amsmath,graphics,longtable,lscape,threeparttable}
\usepackage{amsmath}
\usepackage{url}
\usepackage{pdflscape}
\slugcomment{Draft version}

\shorttitle{A type II Supernova Hubble diagram from the CSP-I, SDSS-II and SNLS surveys.}
\shortauthors{de Jaeger et al.}

\begin{document}

\title{A type II Supernova Hubble diagram from the CSP-I, SDSS-II and SNLS surveys.\altaffilmark{*}} 

\altaffiltext{*}{This paper includes data gathered with the 6.5 m Magellan Telescopes, with the du-Pont and Swope telescopes located at Las Campanas Observatory, Chile; and the Gemini Observatory,Cerro Pachon, Chile (Gemini Program N-2005A-Q-11, GN-2005B-Q-7, GN-2006A-Q-7, GS-2005A-Q-11 and GS-2005B-Q-6, GS-2008B-Q-56). Based on observations collected at the European Organisation for Astronomical Research in the Southern Hemisphere, Chile (ESO Programmes 076.A-0156,078.D-0048, 080.A-0516, and 082.A-0526).}

\author{T. \rm{de} Jaeger$^{1,2,3}$, S. Gonz\'alez-Gait\'an$^{4,1}$, M. Hamuy$^{2,1}$, L. Galbany$^{5}$, J. P. Anderson$^{6}$, M. M. Phillips$^{7}$, M. D. Stritzinger$^{8}$, R. G. Carlberg$^{9}$, M. Sullivan$^{10}$, C. P. Guti\'errez$^{1,2,6}$, I. M. Hook$^{11}$, D. Andrew Howell$^{12,13}$, E. Y. Hsiao$^{7,8,14}$, H. Kuncarayakti$^{1,2}$, V. Ruhlmann-Kleider$^{15}$, G. Folatelli$^{16}$, C. Pritchet$^{17}$, S. Basa $^{18}$}

\affil{%
  (1) Millennium Institute of Astrophysics, Santiago, Chile; dthomas@das.uchile.cl\\
  (2) Departamento de Astronom\'ia - Universidad de Chile, Camino el Observatorio 1515, Santiago, Chile.\\
  (3) Department of Astronomy, University of California, Berkeley, CA 94720-3411, USA; tdejaeger@berkeley.edu \\
  (4) Center for Mathematical Modelling, University of Chile, Beaucheff 851, Santiago, Chile.\\
  (5) PITT PACC, Department of Physics and Astronomy, University of Pittsburgh, Pittsburgh, PA 15260, USA.\\
  (6) European Southern Observatory, Alonso de C\'ordova 3107, Casilla 19, Santiago.\\
  (7) Las Campanas Observatory, Carnegie Observatories, Casilla 601, La Serena, Chile.\\
  (8) Department of Physics and Astronomy, Aarhus University, Ny Munkegade 120, DK-8000 Aarhus C, Denmark.\\
  (9) Department of Astronomy and Astrophysics, University of Toronto, 50 St. george Street, Toronto, ON, M5S 3H4, Canada.\\
  (10) Department of Physics and Astronomy, University of Southampton, Southampton, SO17 1BJ, UK.\\
  (11) Physics Department, Lancaster University, Lancaster LA1 4YB, UK.\\
  (12) Las Cumbres Observatory Global Telescope Network, 6740 Cortona Dr, Suite 102, Goleta, CA 93117, USA \\
  (13) Department of Physics, University of California, Santa Barbara, Broida Hall, Mail Code 9530, Santa Barbara, CA 93106-9530 \\	
  (14) Department of Physics, Florida State University, Tallahassee, FL 32306, USA.\\
  (15) CEA, Centre de Saclay, irfu SPP, 91191 Gif-sur-Yvette, France.\\	
  (16) Instituto de Astrof\'isica de La Plata, CONICET, Paseo del Bosque S/N, B1900FWA, La Plata, Argentina.\\
  (17) University of Victoria, Department of Physics and Astronomy, PO Box 1700, Stn CSC, Victoria, BC, V8W 2Y2, Canada.\\
  (18) Aix Marseille Université, CNRS, LAM (Laboratoire d'Astrophysique de Marseille) UMR 7326, 13388, Marseille, France. \\	
}

\begin{abstract}

The coming era of large photometric wide-field surveys will increase the detection rate of supernovae by orders of magnitude. Such numbers will restrict spectroscopic follow-up in the vast majority of cases, and hence new methods based solely on photometric data must be developed. Here, we construct a complete Hubble diagram of Type II supernovae combining data from three different samples: the Carnegie Supernova Project-I, the Sloan Digital Sky Survey-II SN, and the Supernova Legacy Survey. Applying the Photometric Colour Method (PCM) to 73 Type II supernovae (SNe~II) with a redshift range of 0.01--0.5 and with no spectral information, we derive an intrinsic dispersion of 0.35 mag. A comparison with the Standard Candle Method (SCM) using 61 SNe~II is also performed and an intrinsic dispersion in the Hubble diagram of 0.27 mag is derived, i.e., 13\% in distance uncertainties. Due to the lack of good statistics at higher redshifts for both methods, only weak constraints on the cosmological parameters are obtained. However, assuming a flat Universe and using the PCM, we derive a Universe's matter density: $\Omega_{m}$=0.32$^{+0.30}_{-0.21}$ providing a new independent evidence for dark energy at the level of two sigma.

\end{abstract}

\keywords{cosmology: distance scale -- galaxies: distances and redshifts -- Stars: supernovae: general}

\section{Introduction}

One of the most important investigation in astronomy is to understand the formation and the composition of our Universe. To achieve this goal is very challenging but can be done by measuring distances using astrophysical sources for which the absolute magnitude is known (aka standard candles), and using the Hubble diagram as a classical cosmological test.\\
\indent For more than two decades, Type Ia supernovae (hereafter SNe Ia; \citealt{min41,filippenko97,howell11} and references therein) have been used as standard candles in cosmology (e.g. \citealt{phillips93,hamuy96,riess96,perlmutter97}), and led to the revolutionary discovery of the accelerated expansion of the Universe driven by an unknown force attributed to dark energy \citep{riess98,perlmutter99,schmidt98}. SNe~Ia cosmology today has reached a mature state in which the systematic errors dominate the overall error budget of the cosmological parameters (e.g. \citealt{conley11,rubin13,betoule14}) and further improvement to constrain the nature of the dark energy requires developing as many independent methods as possible.\\
\indent One of the most interesting independent techniques to derive accurate distances and measure cosmological parameters is the use of Type II supernova (hereafter SNe~II)\footnote{Throughout the rest of the text we refer to SNe~II as the two historical groups, SNe~IIP and SNe~IIL, since recent studies showed that SNe~II family forms a continuous class \citep{anderson14a,sanders15,valenti16}. Note that \citet{arcavi13} and \citet{faran14b,faran14a} have argued for two separate populations.}. Even if both SNe~Ia and SNe~II cosmology use in general the same surveys and share some systematic uncertainties like the photometric calibration, other systematic errors are different such as the redshift evolution uncertainties. Furthermore, SNe~II are the result of the same physical mechanism, and their progenitors are better understood than those of SNe Ia \citep{smartt09a}.\\
 
\indent To date several methods have been developed to standardise SNe~II, such as:
\begin{enumerate}
\item{the ``Expanding Photosphere Method'' (EPM) developed by \citet{kirshner74}}
\item{the ``Spectral-fitting Expanding Atmosphere Method'' (SEAM, \citealt{baron04} and updated in \citealt{dessart08})}
\item{the ``Standard Candle Method'' (SCM) introduced by \citet{hamuy02}}
\item{the ``Photospheric Magnitude Method'' (PMM) which is a generalisation of the SCM over various epochs \citep{rodriguez14}}
\item{and the most recent technique the ``Photometric Colour Method'' (PCM; \citealt{dejaeger15b}).}
\end{enumerate}
In this paper, we focus our effort on two different methods: the SCM which is the most common method used to derive SNe~II distances and thus makes easier the comparison with other works, and the PCM being the only purely photometric method in the literature, i.e., which does not require observed spectra. The EPM and the SEAM methods are not discussed in this paper because they require corrections factors computed from model atmospheres (\citealt{eastman96,dessart05}).\\
\indent

The SCM is a powerful method based on both photometric and spectroscopic input parameters which enables a decrease of the scatter in the Hubble diagram from $\sim$ 1 mag to levels of 0.3 mag \citep{hamuy02} and to derive distances with a precision of $\sim $14\%. This method is mainly built on the correlation between the SN~II luminosity and the photospheric expansion velocity 50 days post-explosion. More luminous SNe~II have the hydrogen recombination front at a larger radius and thus, the velocity of the photosphere will be greater in a homologous expansion \citep{kasen09}. Many other works have used an updated version of the SCM where a colour correction is added in order to take into account the host-galaxy extinction. All these studies \citep{nugent06,poznanski09,poznanski10,olivares10,andrea10,dejaeger15b} have confirmed the use of SNe~II as distance indicators finding similar dispersion in the Hubble diagram (0.25-0.30 mag).\\
\indent

Recently, \citet{dejaeger15b} suggested a new method using corrected magnitudes derived only from photometry. In this method, instead of using the photospheric expansion velocity, the standardisation is done using the second, shallower slope in the light curve after maximum, $s_{2}$, which corresponds to the plateau for the SNe~IIP \citep{anderson14a}. {\citet{anderson14a} found that more luminous SNe~II have higher $s_{2}$ (steeper decline, $>$ 1.15 mag per 100 days) confirming previous studies finding that traditional SNe~IIL are more luminous than SNe~IIP \citep{patat94,richardson14}. Using this correlation and adding a colour term, \citet{dejaeger15b} succeeded to reduce the scatter in the low-redshift SNe~II Hubble diagram ($z=$0.01-0.04) to a level of $\sim$ 0.4 mag ($\pm$ 0.05 mag), which corresponds to a precision of 18\% in distances.\\
\indent A better comprehension of our Universe requires the observation of more distant SNe~II. Differences between the expansion histories are extremely small and distinguishing between them will require measurements extending far back in time. The main purpose of the current work is to build a Hubble diagram using the SCM and the PCM as achieved in \citet{dejaeger15b} but adding higher redshift SN samples such as the Sloan Digital Sky Survey II Supernova Survey (SDSS-II SN; \citealt{frieman08,sako14}), and the Supernova Legacy Survey (SNLS; \citealt{astier06,perrett10}).\\
\indent
The paper is organised as follows. In Section 2 we give a description of the data set and Section 3 describes our procedure to perform K-corrections and S-correction, together with line of sight extinction corrections from our own Milky Way. Section 4 presents the Hubble diagram obtained using the PCM while in Section 5 we use the SCM. In Section 6 we discuss our results and conclude with a summary in Section 7.

\section{Data Sample}
In this paper, we use data from three different projects: the Carnegie Supernova Project-I\footnote{\url{http://CSP-I.obs.carnegiescience.edu/}} (CSP-I; \citealt{ham06}), the SDSS-II SN Survey\footnote{http://classic.sdss.org/supernova/aboutsupernova.html} \citep{frieman08}, and the Supernova Legacy Survey\footnote{http://cfht.hawaii.edu/SNLS/} \citep{astier06,perrett10}. These three surveys all used very similar Sloan optical filters permitting a minimisation of the systematic errors. Our sample is listed in Table~\ref{SN_sample}.

\subsection{Carnegie Supernova Project-I}

The CSP-I had guaranteed access to $\sim$ 300 nights per year between 2004-2009 on the Swope 1-m and the du Pont 2.5-m telescopes at the Las Campanas Observatory (LCO), both equipped with high-performance CCD and IR cameras and CCD spectrographs. This observation time allowed the CSP-I to obtain optical-band light-curves 67 SNe~II (with $z \leq 0.04$) with good temporal coverage and more than 500 visual-wavelength spectra for these same objects.\\
\indent The optical photometry ($u$, $g$, $r$, $i$) was obtained after data processing via standard reduction techniques. The final magnitudes were derived relative to local sequence stars and calibrated from observations of standard stars (\citealt{smithja2002}) and are expressed in the natural photometric system of the Swope+CSP-I bands. The spectra were also reduced and calibrated in a standard manner using IRAF.\footnote{IRAF is distributed by the National Optical Astronomy Observatory, which is operated by the Association of Universities for Research in Astronomy (AURA) under cooperative agreement with the National Science Foundation.} A full description can be found in \citet{ham06}, \citet{contreras10}, \citet{stritzinger11}, and \citet{folatelli13}.\\
\indent From the CSP-I sample, we remove six outliers. Three were described in \citet{dejaeger15b} but SN 2005hd has no clear explosion date defined, SN~2008bp is identified as an outlier by \citet{anderson14a}, and SN~2009au was classified at the beginning as a SNe~IIn showing strong interaction. Thus, the total sample used is composed of 61 SNe~II.

\subsection{Sloan Digital Sky Survey-II SN Survey}

\indent The SDSS-II SN Survey was operated during 3-years, from September 2005 to November 2007. Using the 2.5-m telescope at Apache Point Observatory in New Mexico \citep{gunn06}, repeatedly imaged the same region of the sky around the Southern equatorial stripe 82 \citep{stoughton02}. This survey observed about 80 spectroscopically confirmed core-collapse SNe but the main driver of this project was the study of SNe~Ia, involving the acquisition of only one or two spectra per SNe~II.\\
\indent The images were obtained using the wide-field SDSS-II CCD camera \citep{gunn98}, and the photometry was computed using the five $ugriz$ filters defined in \citet{fukugita96}. More information about the data reduction can be found in \citet{york00}, \citet{ivezic04}, and \citet{holtzman08}.\\
\indent A spectroscopic follow up program was performed and uncertainties derived on the redshift measurement are about 0.0005 when the redshift is measured using the host-galaxy spectra, and about 0.005 when the SN spectral features are used.\\
\indent The total SDSS-II SN sample is composed of 16 spectroscopically confirmed SNe~II of which 15 SNe~II are from \citet{andrea10}, and we add one SN~II (SN 2007ny) removed by \citet{andrea10} in his SCM sample due to the absence of explosion date estimation. We derive an explosion date using the \citet{gonzalezgaitan14} rise model. From the SDSS-II sample, we exclude SN~2007nv due to its large $i$-band uncertainties \citep{andrea10}. Note that the majority of spectra were obtained soon after explosion, ans therefore they only exhibit clearly $H_{\alpha}$ $\lambda 6563$ and $H_{\beta}$ $\lambda 4861$ lines but very weak \ion{Fe}{2} $\lambda 5018$ or \ion{Fe}{2} $\lambda 5169$ lines which are often used for the SCM.

\subsection{Supernova Legacy Survey}

In order to obtain a more complete Hubble diagram, we also use higher redshift SNe~II from the SNLS. The SNLS was designed to discover SNe and to obtain a photometric follow-up using the MegaCam imager \citep{boulade03} on the 3.6-m Canada-France-Hawaii Telescope. The observation strategy consisted of obtaining images of the same field every 4 nights during 5 years (between 2003 and 2008), thus, in total more than 470 nights were allocated to this project. Even though the sample was designed for SNe~Ia cosmology and was very successful \citep{guy10,conley11,sullivan11,betoule14}, the observation of many SNe~II with 0.1$\leq$ z $\leq$ 0.5, with good explosion date constraints, and good photometric coverage allowed the use of this sample to previously construct a SN~II Hubble diagram \citep{nugent06}, to constrain SN~II rise-times \citep{gonzalezgaitan14}, and to derive a precise measurement of the core-collapse SN rate \citep{bazin09}.\\
\indent Photometry was obtained in four pass-bands ($g, r, i, z$) similar to those used by the SDSS-II and CSP-I \citep{regnault09}. After each run, the images are pre-processed using the Elixir pipeline \citep{magnier04} and then, sky background subtraction, astrometry, and photometric correction have been performed using two different and independent pipelines. The description of all the data reduction steps can be found in \citet{astier06}, \citet{baumont08}, \citet{regnault09}, \citet{guy10}, \citet{perrett10}, and \citet{conley11}.\\
\indent Due to the redshift of the SNe, and their faintness, spectroscopy was obtained using different large telescopes. All spectra were reduced in a standard way as described in \citet{howell05}, \citet{bronder08}, \citet{ellis08}, \citet{balland09}, and \citet{walker10}.\\
\indent 
We select only SNe~II from the full photometric sample (more than 6000 objects), as achieved by \citet{gonzalezgaitan14} particulary SNe~II with spectroscopic redshift from the SN or the host, a spectroscopic classification or a good photometric classificatio (based on the Gonzalez method described in \citealt{kessler10}), and a well-defined explosion date. The total SNLS sample is composed of 28 SNe~II, 4 of them were used in \citet{nugent06} to derive the first SNe~II high redshift Hubble diagram. For this sample, 16 SNe~II have a spectrum and could potentially be used for the SCM. SN~07D2an was identified as SN~1987A-like event by \citet{gonzalezgaitan14} and is removed from the SNLS sample. As for the SDSS-II sample, the majority of the spectra do not show clear \ion{Fe}{2} $\lambda 5018$ or \ion{Fe}{2} $\lambda 5169$ absorption lines which prevented us to measure the photospheric expansion velocities using these lines. Fortunately, many SNe~II also exhibit a strong $H_{\beta}$ absorption line ($\lambda 4861$) which is useful for the SCM.

\begin{table*}
\centering
\caption{Supernovae sample}
\begin{tabular}[t]{lcccccccccc}
\hline
\hline

SN & AvG & z$_{helio}$ & z$_{CMB}$ (err) &Explosion date & s$_{2}$ & v$_{H\beta}$\tablenotemark{2} & $\mu_{PCM}$ &$\mu_{SCM}$ & Campaign\tablenotemark{3} & Methods \\  & mag &  &  & MJD & mag 100 days$^{-1}$ & km s$^{-1}$ & mag & mag & &\\
\hline

2004er\tablenotemark{1} &0.070 &0.0147 &0.0139 (0.00011) &53271.8(4.0)  &0.41(0.03)      &8100(170)      &33.37(0.06)   &33.84(0.06)  &CSP-I   &PCM/SCM\\
2004fb                  &0.173 &0.0203 &0.0197 (0.00009) &53242.6(4.0)  &0.47(0.10)      &$\cdots$       &$\cdots$      &$\cdots$     &CSP-I   &PCM/SCM\\
2004fc\tablenotemark{1} &0.069 &0.0061 &0.0052 (0.00002) &53293.5(10.0) &$-$0.08(0.03)     &$\cdots$       &$\cdots$      &$\cdots$     &CSP-I   &PCM/SCM\\
2004fx\tablenotemark{1} &0.282 &0.0089 &0.0089 (0.00001) &53303.5(4.0)  &0.70(0.05)      &$\cdots$       &$\cdots$      &$\cdots$     &CSP-I   &PCM/SCM\\
2005J\tablenotemark{1}  &0.075 &0.0139 &0.0151 (0.00001) &53382.7(7.0)  &0.59(0.01)      &6160(160)      &33.84(0.06)   &33.98(0.07)  &CSP-I   &PCM/SCM\\
2005K  			&0.108 &0.0273 &0.0284 (0.00010) &53369.8(7.0)  &1.06(0.08)      &5490(260)      &35.95(0.07)   &35.64(0.10)  &CSP-I   &PCM/SCM\\
2005Z\tablenotemark{1}  &0.076 &0.0192 &0.0203 (0.00003) &53396.7(8.0)  &1.30(0.02)      &7510(160)     &34.42(0.06)   &34.62(0.07)  &CSP-I   &PCM/SCM\\
2005af                  &0.484 &0.0019 &0.0027 (0.00001) &53323.8(15.0) &$-$0.03(0.13)     &$\cdots$       &$\cdots$      &$\cdots$     &CSP-I   &PCM/SCM\\
2005an\tablenotemark{1} &0.262 &0.0107 &0.0118 (0.00010) &53426.7(4.0)  &1.82(0.05)      &6650(170)      &34.02(0.07)   &33.88(0.07)  &CSP-I   &PCM/SCM\\
2005dk\tablenotemark{1} &0.134 &0.0157 &0.0154 (0.00008) &53599.5(6.0)  &0.86(0.03)      &6900(170)      &33.74(0.06)   &33.94(0.07)  &CSP-I   &PCM/SCM\\
2005dn\tablenotemark{1} &0.140 &0.0094 &0.0090 (0.00006) &53601.5(6.0)  &1.22(0.04)      &$\cdots$       &$\cdots$      &$\cdots$     &CSP-I   &PCM/SCM\\
2005dt                  &0.079 &0.0256 &0.0570 (0.00008) &53605.6(9.0)  &$-$0.01(0.07)     &4500(190)      &35.20(0.06)   &35.03(0.08)  &CSP-I   &PCM/SCM\\
2005dw\tablenotemark{1} &0.062 &0.0176 &0.0166 (0.00007) &53603.6(9.0)  &0.76(0.03)      &5040(310)      &34.70(0.07)   &34.41(0.11)  &CSP-I   &PCM/SCM\\
2005dx\tablenotemark{1} &0.066 &0.0267 &0.0264 (0.00010) &53615.9(7.0)  &0.58(0.04)      &4660(190)      &35.80(0.07)   &35.47(0.08)  &CSP-I   &PCM/SCM\\
2005dz\tablenotemark{1} &0.223 &0.0190 &0.0177 (0.00009) &53619.5(4.0)  &0.33(0.02)      &5965(170)      &34.69(0.06)   &34.73(0.07)  &CSP-I   &PCM/SCM\\
2005es\tablenotemark{1} &0.228 &0.0376 &0.0364 (0.00018) &53638.7(10.0) &$\cdots$        &5055(190)      &$\cdots$      &35.64(0.08)  &CSP-I   &PCM/SCM\\
2005gk\tablenotemark{1} &0.154 &0.0292 &0.0286 (0.00010) &53647.8(1.0)  &0.62(0.06)      &$\cdots$       &35.41(0.06)   &$\cdots$     &CSP-I   &PCM/SCM\\
2005lw\tablenotemark{1} &0.135 &0.0257 &0.0269 (0.00013) &53716.8(5.0)  &1.27(0.05)      &7375(180)      &34.88(0.07)   &35.01(0.07)  &CSP-I   &PCM/SCM\\
2005me                  &0.070 &0.0224 &0.0218 (0.00005) &53721.6(6.0)  &0.99(0.10)      &5700(240)      &35.41(0.07)   &35.23(0.08)  &CSP-I   &PCM/SCM\\
2006Y\tablenotemark{1}  &0.354 &0.0336 &0.0341 (0.00010) &53766.5(4.0)  &0.90(0.14)      &6890(160)      &35.73(0.08)   &35.92(0.07)  &CSP-I   &PCM/SCM\\
2006ai\tablenotemark{1} &0.347 &0.0152 &0.0154 (0.00010) &53781.8(5.0)  &1.15(0.05)      &6430(160)      &33.89(0.06)   &33.87(0.06)  &CSP-I   &PCM/SCM\\
2006bc\tablenotemark{1} &0.562 &0.0045 &0.0049 (0.00004) &53815.5(4.0)  &$\cdots$        &$\cdots$       &$\cdots$      &$\cdots$     &CSP-I   &PCM/SCM\\
2006be\tablenotemark{1} &0.080 &0.0071 &0.0075 (0.00003) &53805.8(6.0)  &0.11(0.03)      &$\cdots$       &$\cdots$      &$\cdots$     &CSP-I   &PCM/SCM\\
2006bl\tablenotemark{1} &0.144 &0.0324 &0.0328 (0.00016) &53823.8(6.0)  &1.13(0.09)      &6550(180)      &35.06(0.07)   &35.30(0.07)  &CSP-I   &PCM/SCM\\
2006ee\tablenotemark{1} &0.167 &0.0154 &0.0145 (0.00009) &53961.8(4.0)  &$-$0.61(0.06)     &3330(170)      &33.96(0.06)   &33.54(0.09)  &CSP-I   &PCM/SCM\\
2006it\tablenotemark{1} &0.273 &0.0155 &0.0145 (0.00007) &54006.5(3.0)  &$\cdots$        &$\cdots$       &$\cdots$      &$\cdots$     &CSP-I   &PCM/SCM\\
2006ms\tablenotemark{1} &0.095 &0.0151 &0.0147 (0.00007) &54034.0(12.0) &$-$0.53(0.06)     &$\cdots$       &34.23(0.06)   &$\cdots$     &CSP-I   &PCM/SCM\\
2006qr\tablenotemark{1} &0.126 &0.0145 &0.0155 (0.00007) &54062.8(7.0)  &0.44(0.02)      &5150(160)      &34.82(0.05)   &34.67(0.07)  &CSP-I   &PCM/SCM\\
2007P\tablenotemark{1}  &0.111 &0.0407 &0.0419 (0.00010) &54118.7(3.0)  &0.28(0.12)      &5710(220)      &35.84(0.08)   &35.87(0.08)  &CSP-I   &PCM/SCM\\
2007U\tablenotemark{1}  &0.145 &0.0260 &0.0260 (0.00003) &54134.6(6.0)  &1.40(0.04)      &7050(170)      &34.91(0.06)   &34.96(0.09)  &CSP-I   &PCM/SCM\\
2007W\tablenotemark{1}  &0.141 &0.0097 &0.0107 (0.00007) &54136.8(7.0)  &$-$0.69(0.06)     &3270(160)      &33.78(0.06)   &33.79(0.06)  &CSP-I   &PCM/SCM\\
2007aa\tablenotemark{1} &0.072 &0.0049 &0.0061 (0.00008) &54135.8(5.0)  &$-$0.09(0.09)     &$\cdots$       &$\cdots$      &$\cdots$     &CSP-I   &PCM/SCM\\
2007ab\tablenotemark{1} &0.730 &0.0235 &0.0236 (0.00004) &54123.8(6.0)  &2.87(0.05)      &8580(190)      &35.30(0.06)   &35.08(0.07)  &CSP-I   &PCM/SCM\\
2007av\tablenotemark{1} &0.099 &0.0046 &0.0058 (0.00008) &54175.7(5.0)  &0.22(0.04)      &$\cdots$       &$\cdots$      &$\cdots$     &CSP-I   &PCM/SCM\\
2007hm\tablenotemark{1} &0.172 &0.0251 &0.0241 (0.00008) &54335.6(6.0)  &1.30(0.04)      &6260(200)      &35.92(0.06)   &35.74(0.07)  &CSP-I   &PCM/SCM\\
2007il\tablenotemark{1} &0.129 &0.0215 &0.0205 (0.00008) &54349.8(4.0)  &$-$0.42(0.03)     &6110(160)      &34.81(0.06)   &34.66(0.08)  &CSP-I   &PCM/SCM\\
2007it                  &0.316 &0.0040 &0.0047 (0.00050) &54348.5(1.0)  &$\cdots$        &$\cdots$       &$\cdots$      &$\cdots$     &CSP-I   &PCM/SCM\\
2007oc\tablenotemark{1} &0.061 &0.0048 &0.0039 (0.00007) &54388.5(3.0)  &1.31(0.03)      &$\cdots$       &$\cdots$      &$\cdots$     &CSP-I   &PCM/SCM\\
2007od\tablenotemark{1} &0.100 &0.0058 &0.0045 (0.00006) &54402.6(5.0)  &0.70(0.06)      &$\cdots$       &$\cdots$      &$\cdots$     &CSP-I   &PCM/SCM\\
2007sq\tablenotemark{1} &0.567 &0.0153 &0.0162 (0.00007) &54421.8(3.0)  &0.34(0.05)      &7500(170)      &34.28(0.06)   &34.69(0.06)  &CSP-I   &PCM/SCM\\
2008F\tablenotemark{1}  &0.135 &0.0183 &0.0177 (0.00008) &54470.6(6.0)  &$-$0.63(0.08)     &4825(170)      &34.77(0.07)   &34.88(0.08)  &CSP-I   &PCM/SCM\\
2008M\tablenotemark{1}  &0.124 &0.0076 &0.0079 (0.00002) &54471.7(9.0)  &0.23(0.05)      &$\cdots$       &$\cdots$      &$\cdots$     &CSP-I   &PCM/SCM\\
2008W\tablenotemark{1}  &0.267 &0.0192 &0.0201 (0.00016) &54485.8(6.0)  &0.26(0.04)      &5760(160)      &34.92(0.06)   &34.65(0.07)  &CSP-I   &PCM/SCM\\
2008ag\tablenotemark{1} &0.229 &0.0148 &0.0147 (0.00002) &54479.8(6.0)  &$-$0.32(0.03)     &4760(160)      &33.42(0.06)   &33.42(0.07)  &CSP-I   &PCM/SCM\\
2008aw\tablenotemark{1} &0.111 &0.0104 &0.0114 (0.00007) &54517.8(10.0) &1.49(0.05)      &6900(160)      &33.22(0.06)   &33.19(0.06)  &CSP-I   &PCM/SCM\\
2008bh\tablenotemark{1} &0.060 &0.0145 &0.0154 (0.00007) &54543.5(5.0)  &0.65(0.04)      &6470(180)      &34.55(0.06)   &34.67(0.07)  &CSP-I   &PCM/SCM\\
2008bk\tablenotemark{1} &0.054 &0.0008 &$-$0.0001 (0.00006) &54542.9(6.0) &$\cdots$        &$\cdots$       &$\cdots$      &$\cdots$     &CSP-I   &PCM/SCM\\
2008br                  &0.255 &0.0101 &0.0111 (0.00007) &54555.7(9.0)  &$-$0.65(0.05)     &2420(180)      &34.19(0.06)   &33.33(0.12)  &CSP-I   &PCM/SCM\\
2008bu\tablenotemark{1} &1.149 &0.0221 &0.0222 (0.00003) &54566.8(5.0)  &$-$0.07(0.33)     &5930(200)      &34.59(0.14)   &34.96(0.09)  &CSP-I   &PCM/SCM\\
2008ga\tablenotemark{1} &1.865 &0.0155 &0.0153 (0.00001) &54711.8(4.0)  &1.01(0.05)      &5550(270)      &34.19(0.07)   &34.03(0.09)  &CSP-I   &PCM/SCM\\
2008gi\tablenotemark{1} &0.181 &0.0244 &0.0237 (0.00012) &54742.7(9.0)  &1.31(0.05)      &6420(220)      &34.97(0.07)   &34.95(0.07)  &CSP-I   &PCM/SCM\\
2008gr\tablenotemark{1} &0.039 &0.0229 &0.0218 (0.00015) &54766.5(4.0)  &0.91(0.06)      &7540(170)      &34.42(0.06)   &34.73(0.06)  &CSP-I   &PCM/SCM\\
2008hg\tablenotemark{1} &0.050 &0.0190 &0.0182 (0.00006) &54779.7(5.0)  &$-$1.32(0.56)     &4300(180)      &34.70(0.20)   &34.82(0.08)  &CSP-I   &PCM/SCM\\
2008ho                  &0.052 &0.0103 &0.0096 (0.00050) &54792.7(5.0)  &$-$0.72(0.07)     &$\cdots$       &$\cdots$      &$\cdots$     &CSP-I   &PCM/SCM\\
2008if                  &0.090 &0.0115 &0.0125 (0.00008) &54807.8(5.0)  &1.27(0.04)      &7100(160)      &33.61(0.06)   &33.74(0.06)  &CSP-I   &PCM/SCM\\
2008il                  &0.045 &0.0210 &0.0203 (0.00005) &54825.6(3.0)  &$-$0.07(0.07)     &$\cdots$       &34.90(0.07)   &$\cdots$     &CSP-I   &PCM/SCM\\
2008in                  &0.061 &0.0050 &0.0064 (0.00008) &54822.8(6.0)  &0.03(0.04)      &$\cdots$       &$\cdots$      &$\cdots$     &CSP-I   &PCM/SCM\\
2009N\tablenotemark{1}  &0.057 &0.0034 &0.0046 (0.00008) &54846.7(5.0)  &$-$0.73(0.03)     &$\cdots$       &$\cdots$      &$\cdots$     &CSP-I   &PCM/SCM\\
2009ao\tablenotemark{1} &0.106 &0.0111 &0.0122 (0.00007) &54890.6(4.0)  &0.93(0.10)      &5570(160)      &33.61(0.07)   &33.42(0.07)  &CSP-I   &PCM/SCM\\
2009bu\tablenotemark{1} &0.070 &0.0116 &0.0112 (0.00004) &54907.9(6.0)  &$-$0.31(0.03)     &5670(160)      &33.40(0.07)   &33.61(0.07)  &CSP-I   &PCM/SCM\\
2009bz\tablenotemark{1} &0.110 &0.0108 &0.0113 (0.00004) &54915.8(4.0)  &0.07(0.04)      &6030(160)      &33.67(0.08)   &33.92(0.07)  &CSP-I   &PCM/SCM\\

\hline
\hline
\end{tabular}
\end{table*}

\setcounter{table}{0}
\begin{table*}
\centering
\caption{Supernovae sample--\textit{Continued}}
\begin{tabular}[t]{lcccccccccc}
\hline
\hline

SN & AvG & z$_{helio}$ & z$_{CMB}$ (err) &Explosion date & s$_{2}$ & v$_{H\beta}$\tablenotemark{2} & $\mu_{PCM}$ &$\mu_{SCM}$ & Campaign\tablenotemark{3} & Methods \\  & mag &  &  & MJD & mag 100 days$^{-1}$ & km s$^{-1}$ & mag & mag & &\\
\hline

18321                    &0.080 &0.1041 &0.1065 (0.00050) &54353.6(5.0)  &1.34(0.20)      &6690(180)      &38.09(0.28)   &38.93(0.23)  &SDSS-II  &PCM/SCM\\
2006gq                  &0.096 &0.0697 &0.0687 (0.00050) &53992.4(3.0)  &0.28(0.07)      &4890(230)      &37.34(0.10)   &37.56(0.12)  &SDSS-II  &PCM/SCM\\
2006iw                  &0.137 &0.0308 &0.0295 (0.00050) &54010.7(1.0)  &0.14(0.05)      &6900(510)      &35.44(0.08)   &35.70(0.12)  &SDSS-II  &PCM/SCM\\
2006jl                  &0.504 &0.0555 &0.0554 (0.00050) &54006.8(15.0) &0.52(0.07)      &8640(700)      &$\cdots$      &36.81(0.15)  &SDSS-II  &PCM/SCM\\
2006kn                  &0.194 &0.1199 &0.1190 (0.00050) &54007.0(1.5)  &1.55(0.36)      &6330(200)      &38.43(0.29)   &38.48(0.25)  &SDSS-II  &PCM/SCM\\
2006kv                  &0.080 &0.0619 &0.0620 (0.00050) &54016.5(4.0)  &1.60(0.23)      &5590(190)      &37.26(0.14)   &37.08(0.14)  &SDSS-II  &PCM/SCM\\
2007kw                  &0.074 &0.0681 &0.0672 (0.00050) &54361.6(2.5)  &1.00(0.08)      &6610(250)      &37.02(0.09)   &37.10(0.09)  &SDSS-II  &PCM/SCM\\
2007ky                  &0.105 &0.0737 &0.0727 (0.00050) &54363.5(3.0)  &0.38(0.37)      &5170(170)      &37.62(0.16)   &37.17(0.14)  &SDSS-II  &PCM/SCM\\
2007kz                  &0.320 &0.1275 &0.1270 (0.00050) &54362.6(3.5)  &1.44(0.33)      &6060(200)      &38.71(0.25)   &39.02(0.33)  &SDSS-II  &PCM/SCM\\
2007lb                  &0.496 &0.0379 &0.0375 (0.00050) &54368.8(7.0)  &0.22(0.07)      &7350(370)      &35.31(0.07)   &35.77(0.10)  &SDSS-II  &PCM/SCM\\
2007ld                  &0.255 &0.0270 &0.0267 (0.00500) &54369.6(5.5)  &0.53(0.05)      &6620(420)      &35.07(0.07)   &35.24(0.42)  &SDSS-II  &PCM/SCM\\
2007lj                  &0.118 &0.0500 &0.0489 (0.00500) &54370.2(3.5)  &0.84(0.06)      &5610(300)      &36.72(0.07)   &36.43(0.25)  &SDSS-II  &PCM/SCM\\
2007lx                  &0.120 &0.0571 &0.0559 (0.00050) &54374.5(8.0)  &0.47(0.11)      &5520(200)      &37.07(0.22)   &37.10(0.11)  &SDSS-II  &PCM/SCM\\
2007nr                  &0.079 &0.1400 &0.1389 (0.00050) &54353.5(5.0)  &1.25(0.46)      &5230(190)      &39.53(0.40)   &39.27(0.30)  &SDSS-II  &PCM/SCM\\
2007nw 			&0.204 &0.0571 &0.0555 (0.00050) &54372.2(7.0)  &$-$0.06(0.17)     &5810(200)      &37.17(0.12)   &36.96(0.12)  &SDSS-II  &PCM/SCM\\
2007ny 			&0.080 &0.1429 &0.1419 (0.00050) &54367.7(7.0)  &1.39(1.44)      &6860(190)      &39.43(0.64)   &39.14(0.45)  &SDSS-II  &PCM/SCM\\
03D1bo 			&0.066 &0.3279 &0.3271 (0.00100) &52888.0(4.0)  &$\cdots$        &$\cdots$       &$\cdots$      &$\cdots$     &SNLS  &PCM/SCM\\
03D1bz 			&0.066 &0.2939 &0.2931 (0.00100) &52914.0(4.0)  &0.16(0.23)      &$\cdots$       &$\cdots$      &$\cdots$     &SNLS  &PCM/SCM\\
03D3ce 			&0.026 &0.2880 &0.2884 (0.00100) &52780.0(10.0) &$\cdots$        &$\cdots$       &$\cdots$      &$\cdots$     &SNLS  &PCM/SCM\\
03D4az			&0.076 &0.4079 &0.4069 (0.00100) &52808.0(4.0)  &$-$0.47(1.49)     &$\cdots$       &$\cdots$      &$\cdots$     &SNLS  &PCM\\
03D4bl			&0.072 &0.3179 &0.3169 (0.00100) &52822.0(3.0)  &1.32(1.46)      &$\cdots$       &41.43(0.57)   &$\cdots$     &SNLS  &PCM\\
03D4da 			&0.078 &0.3279 &0.3269 (0.00100) &52874.0(7.0)  &$\cdots$        &$\cdots$       &$\cdots$      &$\cdots$     &SNLS  &PCM/SCM\\
04D1ha			&0.073 &0.4839 &0.4831 (0.00100) &53233.0(3.0)  &0.11(0.42)      &$\cdots$       &42.13(0.44)   &$\cdots$     &SNLS  &PCM\\
04D1ln 			&0.071 &0.2069 &0.2062 (0.00100) &53274.0(5.0)  &0.32(0.12)      &$\cdots$       &40.28(0.11)   &$\cdots$     &SNLS  &PCM/SCM\\
04D1nz			&0.072 &0.2629 &0.2621 (0.00100) &53264.0(4.0)  &0.13(0.37)      &$\cdots$       &40.74(0.30)   &$\cdots$     &SNLS  &PCM\\
04D1pj 			&0.076 &0.1559 &0.1552 (0.00100) &53304.0(8.0)  &0.02(0.14)      &5975(230)      &39.13(0.09)   &39.37(0.10)  &SNLS  &PCM/SCM\\
04D1qa			&0.072 &0.1719 &0.1711 (0.00100) &53300.0(3.0)  &$-$0.10(0.40)     &$\cdots$       &39.65(0.19)   &$\cdots$     &SNLS  &PCM\\
04D2dc 			&0.053 &0.1849 &0.1861 (0.00100) &53040.0(25.0) &0.14(0.19)      &$\cdots$       &$\cdots$      &$\cdots$     &SNLS  &PCM/SCM\\
04D4fu 			&0.072 &0.1329 &0.1319 (0.00100) &53213.0(6.0)  &0.23(0.60)      &4785(200)      &39.21(0.23)   &39.02(0.10)  &SNLS  &PCM/SCM\\
04D4hg 			&0.073 &0.5169 &0.5159 (0.00100) &53233.0(3.0)  &$\cdots$        &$\cdots$       &$\cdots$      &$\cdots$     &SNLS  &PCM\\
05D1je 			&0.071 &0.3089 &0.3081 (0.00100) &53647.0(5.0)  &$-$0.22(0.80)     &$\cdots$       &41.46(0.36)   &$\cdots$     &SNLS  &PCM\\
05D2ai 			&0.052 &0.2489 &0.2501 (0.00100) &53377.0(9.0)  &$\cdots$        &$\cdots$       &$\cdots$      &$\cdots$     &SNLS  &PCM/SCM\\
05D2ed 			&0.051 &0.1959 &0.1971 (0.00100) &53417.0(5.0)  &$-$0.08(0.34)     &$\cdots$       &39.50(0.16)   &$\cdots$     &SNLS  &PCM\\
05D2js 			&0.051 &0.0926 &0.0934 (0.00100) &53670.0(17.0) &$\cdots$        &$\cdots$       &$\cdots$      &$\cdots$     &SNLS  &PCM/SCM\\
05D2or 			&0.051 &0.2470 &0.2480 (0.00100) &53731.0(3.0)  &0.77(0.46)      &$\cdots$       &$\cdots$      &$\cdots$     &SNLS  &PCM\\
05D4ar 			&0.072 &0.1909 &0.1889 (0.00100) &53520.0(25.0) &0.80(0.50)      &$\cdots$       &40.16(0.24)   &$\cdots$     &SNLS  &PCM/SCM\\
05D4cb 			&0.073 &0.1999 &0.1989 (0.00100) &53563.0(3.0)  &0.41(0.13)      &$\cdots$       &39.66(0.11)   &$\cdots$     &SNLS  &PCM\\
05D4dn 			&0.073 &0.1909 &0.1889 (0.00100) &53605.0(7.0)  &0.55(0.42)      &4970(210)      &40.19(0.23)   &39.90(0.22)  &SNLS  &PCM/SCM\\
05D4du 			&0.072 &0.3099 &0.3089 (0.00100) &53585.0(5.0)  &0.01(0.30)      &$\cdots$       &41.15(0.25)   &$\cdots$     &SNLS  &PCM\\
06D1jx 			&0.079 &0.1349 &0.1342 (0.00100) &54068.0(6.0)  &$-$0.39(0.14)     &6110(190)      &38.82(0.08)   &39.23(0.10)  &SNLS  &PCM/SCM\\
06D2bt 			&0.051 &0.0779 &0.0791 (0.00100) &53745.0(10.0) &$-$0.02(0.23)     &5965(200)      &37.69(0.10)   &37.90(0.08)  &SNLS  &PCM/SCM\\
06D2ci 			&0.053 &0.2199 &0.2211 (0.00100) &53768.0(4.0)  &1.08(0.25)      &$\cdots$       &40.46(0.16)   &$\cdots$     &SNLS  &PCM\\
06D3fr 			&0.025 &0.2749 &0.2754 (0.00100) &53883.0(4.0)  &$\cdots$        &$\cdots$       &$\cdots$      &$\cdots$     &SNLS  &PCM/SCM\\
06D3gg 			&0.024 &0.2659 &0.2663 (0.00100) &53897.0(6.0)  &$\cdots$        &$\cdots$       &$\cdots$      &$\cdots$     &SNLS  &PCM/SCM\\

\hline
\hline
\end{tabular}
\begin{tablenotes}
\item$^1$ SNe~II used in \citet{dejaeger15b}.
\item$^2$ 45 days post-explosion.
\item$^3$ CSP-I=Carnegie Supernova Project-I, SDSS-II=Sloan Digital Sky Survey II SN, SNLS= Supernova Legacy Survey.
\end{tablenotes}
\tablecomments{In the first column the SN name, followed by its reddening due to dust in our Galaxy \citep{schlafly11} are listed. In column 3, we list host-galaxy heliocentric recession velocities. These are taken from the NASA Extragalactic Database (NED: \url{http://ned.ipac.caltech.edu/}). In column 4, we list the host-galaxy velocity in the CMB frame using the CMB dipole model presented by \citet{fixsen96}. In column 5, the explosion epoch is presented. In column 6, the s$_{2}$ value in the $i$ band (defined in Section \ref{PCM}) is listed followed by the H$\beta$ velocity at an epoch of 45 days after the explosion (see Section \ref{SCM}) in column 7. In columns 8 and 9 we present the distance modulus measured using PCM and SCM respectively. In column 10, we list the survey from which the SN~II originates, and finally in the last column we show for each the SN the method available, i.e, whether a spectrum is available or not.}
\label{SN_sample}
\end{table*}

\section{Methodology}\label{methodology}

\subsection{Background}

The photon flux observed in one photometric system is affected by four different sources: the dust in our Milky Way (AvG), the expansion of the Universe (K-correction; \citealt{oke68,hamuy93,kim96,nugent02}), the host-galaxy extinction (Avh), and by the difference between the natural photometric system used to obtain observations and the standard photometric system (S-correction; \citealt{stritzinger02}). In the following we describe how we account for these factors and place all of the photometry on a common photometric system.\\
\indent
Correcting for AvG is straightforward using the value derived by \citet{schlafly11} and gathered in the NASA/IPAC Extragalactic Database (NED\footnote{NASA/IPAC Extragalactic Database (NED) is operated by the Jet Propulsion Laboratory, California Institute of Technology, under contract with the National Aeronautics and Space Administration.}). Correcting for Avh is more complicated. To date, no accurate methods exist. For example, the equivalent width of the \ion{Na}{1} doublet lines \citep{tur03} is a bad proxy for extinction \citep{poznanski11}. For the colour-colour diagram, and multi-band fit methods \citep{folatelli13,rodriguez14} a better understanding and estimation of the SNe~II intrinsic colour is necessary to derive a good approximation of Avh. To take into account the host-galaxy extinction, similarly to what it has been done in SNe~Ia cosmology, we add a colour term correction in the standardisation method which takes care of the magnitude-colour variations independently of their origin. The AvG, K-, and S-corrections (AKS) are finally simultaneously computed using the cross-filter K-corrections defined by \citet{kim96} from an observed filter (CSP-I, SDSS-II or SNLS) and the standard system. The CSP-I natural system was chosen as the common photometric system and thus, we transform all the photometry to the CSP-I system. The AKS correction depends on anything that could affect the SED (Spectral Energy Distribution) continuum; it is thus very important to adjust the continuum to have the same colour as the SN~II through a color-matching function \citep{nugent02,hsiao07}. In Section \ref{procedure} we describe the procedure to apply the AKS correction.

\subsection{Procedure}\label{procedure}

In practice, to apply the AKS correction a SED template series is needed. We adopt a sequence of theoretical spectral models from \citet{dessart13} consisting of a SN progenitor with a main-sequence mass of 15 ${\rm M}_{\odot}$, solar metallicity $Z=$0.02, zero rotation and a mixing-length parameter of 3. The choice of the model was motivated by the very good match of the theoretical model to the data of the prototypical SNe~II (SN~1999em). We describe step by step the method to transform an observed magnitude to the CSP-I photometric system. The two first steps can be found in \citet{dejaeger15b} and consist of selecting the theoretical spectrum that has an epoch closest in time to the respective epoch of the observed light-curve point of the SN in consideration and bringing it to the observed frame using the heliocentric redshift.

\begin{enumerate}

\item{We adjust the SED continuum to match the observed colour by comparing synthetic magnitudes with observed magnitudes using the zero points defined in \citet{fukugita96}, \citet{stritzinger02}, and \citet{regnault09}. A color-matching (CM) function $CM(\lambda)$ is obtained and used to correct our model spectrum. Finally, we obtain $f^{obs}_{CM}({\lambda})=CM(\lambda) \times f^{obs}({\lambda})$ and calculate the magnitude in the observer's frame :

\begin{equation}
m_{Y}=-2.5log_{10}\left[\int f^{obs}_{CM}(\lambda)S_{Y}\lambda d\lambda\right]+ZP_{Y},
\end{equation}

with $\lambda$ the wavelength, $S_{Y}$ the transmission function of filter $Y$ (CSP-I, SDSS-II, or SNLS) 
and $ZP_{Y}$ the zero point of filter $Y$.}

\item{Using a \citet{car89} law, we correct for AvG and bring the unredenned spectrum to the rest-frame $f^{rest,AvG}_{CM}({\lambda^{'}})$. The magnitude X in the rest-frame is computed :

\begin{equation}
m^{CSP-I}_{X}=-2.5log_{10}~\int {f^{rest,AvG}_{CM}(\lambda)S^{CSP-I}_{X}\lambda d\lambda}+ZP^{CSP-I}_{X},
\end{equation}

where $S^{CSP-I}_{X}$ is the transmission function of filter $X$ and $ZP^{CSP-I}_{X}$ is the zero point of filter $X$ of the CSP-I system (see \citealt{contreras10,stritzinger11}). The choice of the filter X depends on the redshift. At low redshift the K-correction is not so important, so X and Y are the same bands, but at higher redshifts it is not the case. We need to know in which band a photon received in the Y band has been emitted. For this, we calculate the effective wavelength of the filter Y corrected by the (1+z$_{hel}$) factor and select the closest effective wavelength among the CSP-I filters ($u$, $g$, $r$, $i$). Then the AKS is obtained doing:

\begin{equation}
AKS_{XY}=m_{Y}-m^{CSP-I}_{X}.
\end{equation}
}
\end{enumerate}

\indent The AKS corrections are sensitive to the choice of spectral template, thus, in order to estimate the systematic errors, we try another model: Nugent's templates\footnote{\url{https://c3.lbl.gov/nugent/index.html}}. These templates are based on the models from \citet{baron04}. The comparison between Nugent's templates and the Dessart's model \citep{dessart13}, leads to a mean difference of 0.004 mag in $r$ band (with a standard deviation of 0.02 mag) and 0.03 mag (with a standard deviation of $\pm$ 0.06 mag) in $i$ band. It is important to note that, both models were created to fit the same observed data (SN 1999em spectra) and a comparison between the Dessart's model and observed spectra using the CSP-I sample was achieved in \citet{dejaeger15b} who derived a good agreement.

\section{The Photometric colour Method (PCM)}\label{PCM}

\subsection{Methodology}

The basic idea of this method is to correct and standardise the apparent magnitude using two photometric parameters: $s_{2}$ which is the slope of the plateau, and a colour term at a specific epoch \citep{dejaeger15b}. To measure the $s_{2}$ in all the bands we use a Python program, which consists of performing a least-squares fitting of the AKS corrected light-curves, with one and two slopes (sometimes the first decline after the maximum is not visible). To choose between one or two slopes, the statistical method F-test is performed. The slope is measured in each band ($g$, $r$, and $i$) but only the values used in this work ($i$ band) are listed in Table \ref{SN_sample}. The minimisation of intrinsic dispersion in the Hubble diagram is our figure of merit and allows us to find the best combination possible of filter, colour, and epoch as done by \citet{dejaeger15b}. Note that using only the CSP-I sample we derive the $V$-band light curve slopes and perform a sanity check by comparing these values with those found by \citet{anderson14a}. We obtain a very good agreement. Also, using 51 SNe~II, \citet{galbany16a} confirmed the relation found by \citet{anderson14a} between s$_{2}$ and the absolute magnitude in different bands. The observed magnitudes can be modelled as:\\

\begin{equation}
\begin{split}
m^{model}_{\lambda 1}=&M_{\lambda 1}-\alpha  s_{2} + \beta_{\lambda 1}  (m_{\lambda 2}-m_{\lambda 3})\\
 +& 5  log_{10} (d_{L}(z_{CMB}|\Omega_{m},\Omega_{\Lambda}))+25,
\end{split}
\end{equation}

where $(m_{\lambda 2}-m_{\lambda 3})$ is the colour, $d_{L}(z_{CMB}|\Omega_{m},\Omega_{\Lambda})$ is the luminosity distance for a cosmological model depending on: the cosmological parameters $\Omega_{m},\Omega_{\Lambda}$, the CMB redshift $z_{CMB}$, and the Hubble constant. Finally, $\alpha$, $\beta_{\lambda_{1}}$, and $M_{\lambda_{1}}$ are also free parameters with $M_{\lambda_{1}}$ corresponding to the absolute magnitude in the filter $\lambda_{1}$. Note that the s$_{2}$ and the colour distributions are centered, i.e., we use (s$_{2}$-$<s_{2}>$) and $(m_{\lambda 2}-m_{\lambda 3})$-$<(m_{\lambda 2}-m_{\lambda 3})>$ where $<s_{2}>$ and $<(m_{\lambda 2}-m_{\lambda 3})>$ are the mean values of the slope and the colour respectively for the whole sample (CSP-I+SDSS-II+SNLS).\\
\indent Since we do not have in our sample any SN~II with an accurate distance estimation (e.g. from Cepheid measurements), we only measure relative distances and define the ``Hubble Constant free'' absolute magnitude as $\mathcal{M}_{\lambda 1}$=M$_{\lambda 1}$-5 log$_{10}$(H$_{0}$) + 25 and $\mathcal{D}_{L}$=H$_{0}$$d_{L}$ as done in many previous works \citep{perlmutter99,nugent06,poznanski09,andrea10}. The apparent magnitude is finally written as:
\begin{equation}
\begin{split}
m^{model}_{\lambda 1}=&\mathcal{M}_{\lambda 1}-\alpha  s_{2} + \beta_{\lambda 1}  (m_{\lambda 2}-m_{\lambda 3})\\ +& 5 log_{10} (\mathcal{D}_{L}(z_{CMB}|\Omega_{m},\Omega_{\Lambda})).
\end{split}
\label{m_model}
\end{equation}

From this equation, one can derive the $\alpha$, $\beta_{\lambda_{1}}$ and $\mathcal{M}_{\lambda 1}$ and the cosmological parameters $\Omega_{m},\Omega_{\Lambda}$. Due to our imperfect knowledge of SN~II physics, another free parameter named $\sigma_{int}$ needs to be added in order to include the intrinsic scatter not accounted for measurement errors. This dispersion is the minimum statistical uncertainty in any distance determination using PCM.\\

\indent In cosmology, when we compare observations to predictions of a parameter-dependent model, Bayesian inference is the standard procedure. This approach tells us how to update our knowledge from a ``prior'' distribution to a new probability density named ``posterior''. Thus, to define the posterior probability density one needs to define a likelihood function and a prior function. As the likelihood function, we choose that defined by \citet{andrea10}:

\begin{equation}
-2ln(\mathcal{L})=\sum_{SN} \left \{ \frac{\left [m^{obs}_{\lambda 1}- m^{model}_{\lambda 1} \right ]^{2}}{\sigma^{2}_{tot}} +ln(\sigma^{2}_{tot}) \right \},
\label{likelihood}
\end{equation}

where we sum over all SNe~II available for one specific epoch, m$_{\lambda 1}^{obs}$ is the observed magnitude corrected for AKS, m$_{\lambda 1}^{model}$ the model defined in equation \ref{m_model}, and the total uncertainty $\sigma_{tot}$, corresponding to the error propagation of the model, is defined as:

\begin{equation}
\begin{split}
\sigma^{2}_{tot}=&\sigma^{2}_{m_{\lambda 1}} + (\alpha \sigma_{s2})^2 + (\beta \sigma_{(m_{\lambda 2}-m_{\lambda 3})})^2\\&+\left (\sigma_{z} \frac{5(1+z)}{z(1+z/2)ln(10)}\right )^{2}+\sigma^{2}_{int}.
\end{split}
\end{equation}

For the relation between the redshift uncertainty and the associated magnitude uncertainties, we use the empty Universe approximation \citep{conley11}. The second logarithmic term comes from the normalisation of the likelihood function and is useful in order to not obtain large values of $\alpha$, $\beta$, and $\mathcal{M}_{\lambda 1}$ which could be favoured by the first part of the log-likelihood.\\

\indent Our prior probability distribution is defined to have uniform probability for $0 \leq\Omega_{m} \leq 1$ or $\alpha$, $\beta$, $\mathcal{M}_{\lambda 1}$ $\neq$0 but otherwise has zero probability. We also attempted to use a Gaussian prior, however no differences were found in the fit parameters. To explore the posterior probability density, a Monte Carlo Markov Chain (MCMC) simulation is performed. The MCMC calculation is run using a Python package called \textsc{EMCEE} developed by \citet{foreman13} and using 500 walkers and 1000 steps (for the convergence as suggested by the authors of the EMCEE package, we checked the fraction acceptance which should be between 0.2-0.5). The EMCEE package uses an ensemble of walkers which can be moved in parallel and not a single iterative random walker (Goodman-Weare algorithm versus Metropolis-Hastings algorithm).\\
\indent The entire sample available for this method contains 105 SNe~II. From this sample, in order to avoid peculiar galaxy motions, we select only the SNe~II with $cz_{CMB}$ $\geq$ 3000 km s$^{-1}$. After this first cut, the sample size drops to 89 SNe~II including 45 SNe~II from CSP-I, 16 SNe~II from SDSS-II and 28 SNe~II from SNLS.

\subsection{Results}

In this results section, we will first attempt to extend the low-redshift PCM Hubble diagram \citep{dejaeger15b} to higher redshifts, and then investigate whether such efforts can constrain cosmological parameters (Section \ref{omega_m_PCM}).

\subsubsection{Fixed cosmology}\label{pcm_fixed}

In order to test the method at higher redshifts, we first assume a fiducial $\Lambda$CDM model, i.e., a flat Universe ($\Omega_{m}$+$\Omega_{\Lambda}$=1) with $\Omega_{m}=0.3$ and $\Omega_{\Lambda}=0.7$. We assume also a Hubble constant of 70 km s$^{-1}$ Mpc$^{-1}$ because we are not able to derive a Hubble constant using our low-redshift sample (lack of SNe~II with Cepheid measurements).\\
\indent Using all the filter combinations available for the three surveys ($g$, $r$,$i$), we find a minimum intrinsic dispersion for the ($r$-$i$) colour associated with the $i$ band for an epoch of rest-frame day 40. At this specific epoch, only 73 SNe~II have photometric data. In this work we interpolate all the magnitudes and colours at this epoch but we do not extrapolate, leaving us with only 73 SNe~II from the entire sample (89 SNe~II). In Figure \ref{HD_PCM}, we present the final Hubble diagram using the PCM. For the 73 SNe~II in the Hubble flow available at this epoch and at this specific colour, we obtain an intrinsic scatter of 0.35 mag, i.e., 16\% in distance errors. The use of the PCM allows us to reduce the intrinsic scatter from 0.57 mag (raw magnitudes) to 0.35 mag , i.e., an improvement of 10\% in distance errors. This scatter is somewhat lower than that found by \citet{dejaeger15b} due to the higher redshift SNe~II (0.4-0.44 mag), which as it will be shown in Section \ref{com_sample}, exhibit a smaller range in absolute magnitude. The Bayesian inference procedure using the likelihood defined in equation \ref{likelihood} gives $\alpha = 0.36^{+0.06}_{-0.06}$, $\beta = 0.70^{+0.29}_{-0.29} $, $\mathcal{M}_{\lambda 1}=-1.09^{+0.05}_{-0.05}$. Using only the CSP-I sample as done in \citet{dejaeger15b}, we find $\alpha = 0.39^{+0.08}_{-0.08}$, $\beta = 0.80^{+0.47}_{-0.48} $, $\mathcal{M}_{\lambda 1}=-1.06^{+0.06}_{-0.07}$. These values are consistent with those derived by \citet{dejaeger15b}. From $\mathcal{M}_{\lambda 1}$ and with $H_{0}$=70 km s$^{-1}$ Mpc$^{-1}$ an absolute magnitude M$_{i}$ of -16.84$^{+0.06}_{-0.06}$ mag is obtained. This value is relatively low compared to the value reported by \citet{richardson14}, we do not account for the intrinsic colour $(m_{\lambda 2}-m_{\lambda 3})_{int}$, where ``int'' is for intrinsic. Indeed, in our model (equation \ref{m_model}) the host-galaxy extinction is taken into account using the observed colour, however only the excess colour ($E(m_{\lambda 2}-m_{\lambda 3})$) is directly related to the Avh and should be used. In the following equations, we show the relation between the Avh, the excess colour and the intrinsic colour:
\begin{equation}
\begin{split}
A_{\lambda 1}=&\beta_{\lambda 1} \times E(m_{\lambda 2}-m_{\lambda 3})\\&
=\beta_{\lambda 1} \times (m_{\lambda 2}-m_{\lambda 3})-\beta_{\lambda 1}(m_{\lambda 2}-m_{\lambda 3})_{int}\\&
=\beta_{\lambda 1} \times (m_{\lambda 2}-m_{\lambda 3})+constant.
\end{split}
\end{equation}

Thus, the intrinsic colour is degenerate with the $\mathcal{M}_{\lambda 1}$, so any approximation (we assume that the intrinsic colour is zero) on this value has consequences in the absolute magnitude determination.\\

\begin{figure}
\includegraphics[width=9.0cm]{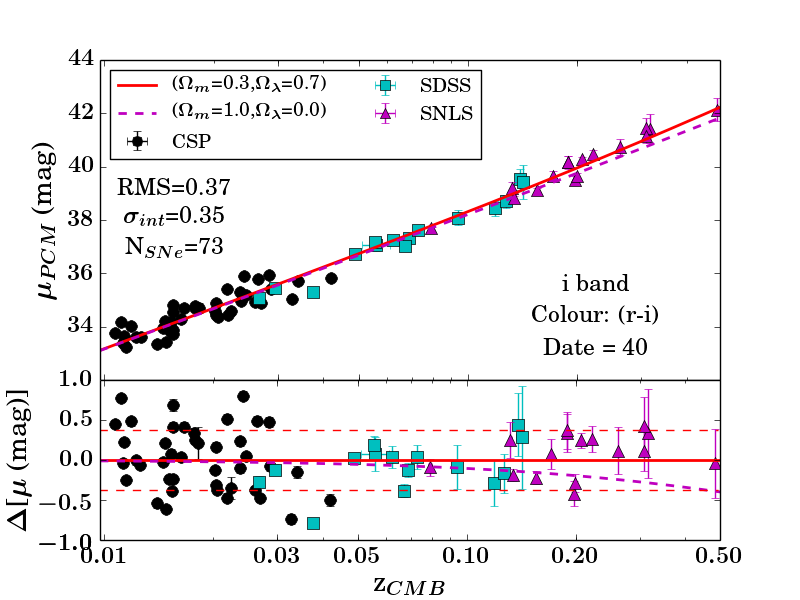}
\caption{Hubble diagram for SNe~II, using the PCM and all the SNe~II available at this epoch from the CSP-I, SDSS-II, and SNLS sample respectively. Black dots represent the SNe~II from the CSP-I whereas the cyan squares and magenta triangles are the SDSS-II and the SNLS sample respectively. The red line is the Hubble diagram for the $\Lambda$CMB ($\Omega_{m}$=0.3 and $\Omega_{\Lambda}$=0.7) and in magenta for an Einstein-de Sitter cosmological model ($\Omega_{m}$=1.0 and $\Omega_{\Lambda}$=0.0). In both models, we assume a Hubble constant of 70 km s$^{-1}$ Mpc$^{-1}$. In the bottom panel, the residuals with respect to the $\Lambda$CMB are shown. We also present the number of SNe~II available at this epoch (N$_{SNe}$), the epoch after the explosion (Date), the Root Mean Square (RMS) and the intrinsic dispersion ($\sigma_{int}$).}
\label{HD_PCM}
\end{figure}

\begin{table*}[ht]
\caption{PCM-fit Parameters.}
\centering
\begin{tabular}{cccccc}
\hline
\hline
Data Set & $\alpha$  & $\beta$ &$M_{i}$ &$\sigma_{int}$ &SNe\\
\hline
CSP-I &0.39$^{+0.08}_{-0.08}$ &0.80$^{+0.47}_{-0.48}$ &$-$16.84$^{+0.06}_{-0.07}$ &0.41$^{+0.05}_{-0.04}$ &42\\
CSP-I$+$SDSS-II &0.38$^{+0.07}_{-0.07}$ &0.75$^{+0.36}_{-0.36}$ &$-$16.87$^{+0.05}_{-0.05}$ &0.38$^{+0.04}_{-0.04}$ &57\\
CSP-I$+$SNLS &0.36$^{+0.07}_{-0.07}$ &0.87$^{+0.34}_{-0.35}$ &$-$16.83$^{+0.05}_{-0.05}$ &0.37$^{+0.04}_{-0.03}$ &58\\
SDSS-II$+$SNLS &0.28$^{+0.10}_{-0.10}$ &0.69$^{+0.36}_{-0.36}$ &$-$16.90$^{+0.06}_{-0.06}$ &0.27$^{+0.05}_{-0.04}$ &31\\
CSP-I$+$SDSS-II$+$SNLS &0.36$^{+0.06}_{-0.06}$ &0.71$^{+0.29}_{-0.28}$&$-$16.85$^{+0.05}_{-0.05}$ & 0.36$^{+0.03}_{-0.03}$ &73\\
\hline
\multicolumn{5}{l}{%
\begin{minipage}{12.5cm}%
Note: Best-fit values and the associated errors for each parameter for different samples using the PCM.%
\end{minipage}%
}\\
\end{tabular}
\label{PCM_param}
\end{table*}

\indent Systematic errors coming from the SN~II sample at different redshift are investigated by looking at the fitting parameters evolution using different samples, i.e., CSP-I, CSP-I+SDSS-II, CSP-I+SNLS, SDSS-II+SNLS, and CSP-I+SDSS-II+SNLS. In Table \ref{PCM_param} a summary of these values is shown. As it is seen from this table, the fitting parameters remain similar within the uncertainties for the different samples which means that there does not seem to be a systematic redshift or SNe~II sample evolution. We can also study how the parameters are affected by photometric errors. If we arbitrarily increase the centred color distribution by an offset (0.01 mag and 0.5 mag) almost all the fitting parameters remain also similar. Only the Hubble constant free absolute magnitude $\mathcal{M}_{\lambda 1}$ changes from to $-$1.09 to $-$1.49 but is explained by the fact that $\mathcal{M}_{\lambda 1}$ and the intrinsic colour are degenerated.\\

\indent A residual analysis between the data and the $\Lambda$CDM cosmological model is also performed by testing for normality (cf. Anderson-Darling test, \citealt{stephen74}), for autocorrelation (Durbin-Watson test, \citealt{durbin50}), stationarity (Kwiatkowski-Phillips-Schmidt-Shin test, \citealt{KPSS92}), and outliers (Chauvenet's criterion, \citealt{chauvenet1863}). For the normality test, at a significance level from 1\% to 15\% we cannot reject the null hypothesis that the residuals come from a Gaussian distribution. In the same way, we cannot reject the null hypothesis of stationarity. Additionally, we do not find existence of autocorrelation and no value should be eliminated according to the Chauvenet's criterion.

\subsubsection{$\Omega_{m}$ derivation} \label{omega_m_PCM}

In this section, we try to put some constraints on the cosmological parameters ($\Omega_{m}$ and $\Omega_{\Lambda}$) assuming a Hubble constant $H_{0}$=70 km s$^{-1}$ Mpc$^{-1}$. Due to the lack of higher redshift SNe~II (only one with z $\geq$ 0.4), it is difficult to differentiate between cosmological models, and so, to derive a meaningful constraint on cosmology. Indeed, keeping $\Omega_{m}$, $\Omega_{\Lambda}$ as free parameters in equation \ref{m_model}, we are not able to obtain constraints with reasonable error bars. So, we assume a flat Universe, i.e., $\Omega_{m}$+$\Omega_{\Lambda}$=1 and leave only $\Omega_{m}$ as a free parameter. Figure \ref{HD_PCM_cosmo} shows a corner plot with all the one and two dimensional projections for the five free parameters $\alpha$, $\beta$, $\mathcal{M}_{\lambda 1}$, $\sigma_{int}$, and $\Omega_{m}$. The four first parameters are defined by a Gaussian distribution (top figure of each column) with small error bars. The values derived are consistent with that found with a fixed cosmology: $\alpha = 0.36^{+0.06}_{-0.06}$, $\beta = 0.71^{+0.29}_{-0.28} $, $\mathcal{M}_{\lambda 1}=-1.08^{+0.05}_{-0.05}$, and $\sigma_{int}=0.36^{+0.03}_{-0.03}$.\\
\indent For the matter density, the distribution does not look like a Gaussian distribution but the distribution width decreases as the matter density value increases. A value for the matter density of $\Omega_{m}= 0.32^{+0.30}_{-0.21}$ is derived which gives a density of dark energy of $\Omega_{\lambda}=0.68^{+0.21}_{-0.30}$. These values are consistent with the $\Lambda$CDM cosmological model with uncertainties far from the precision achieved recently using SNe~Ia, $\Omega_{m}$=0.295 $\pm$ 0.034 (from \citealt{betoule14} with $\sim$740 SNe~Ia up to a redshift of 1.2, see also \citealt{rest14} and \citealt{scolnic14} for other results). However, these errors are comparable to those found by \citet{perlmutter97} for which the authors using $\sim$ 20 SNe~Ia (7 SNe~Ia with $z$ between 0.3-0.5) derived an uncertainty in the matter density $\Delta$$\Omega_{m}$ $\sim$ 0.30. Note that the minimisation of the negative likelihood defined in Equation \ref{likelihood} is found for a value $\Omega_{m}$= 0.17 $\pm$ 0.30 (blue line in Figure \ref{HD_PCM_cosmo}). 
\indent To test the sensitivity of $\Omega_{m}$ and its uncertainty to the systematic errors we double the errors on s$_{2}$. The minimum of the negative log of the likelihood function is obtained for the Universe's matter density 0.22 instead of 0.17. Using the MCMC simulation, we derive $\Omega_{m}= 0.36^{+0.31}_{-0.23}$. For this test the other fitting parameters ($\alpha$, $\beta$, and $\mathcal{M}_{\lambda 1}$) do not show any variation and remain similar within the uncertainties (variation only of $\sim$ 0.04 in average for each parameter).\\

\indent The shallow drop in the matter density distribution (Figure \ref{HD_PCM_cosmo}) and the relatively low intrinsic dispersion in the Hubble diagram obtained are encouraging to derive cosmological parameters with reasonable uncertainties in the future. Indeed, with this method, we can correct the apparent magnitude using solely photometric input and thus add more SNe~II in the Hubble diagram and at higher redshifts for which it is difficult to obtain spectrum with sufficient signal to noise ratio. This work demonstrates how the PCM can be extended to high redshift objects and will be an asset for the next generation of surveys. Even if SNe~Ia offer more precise distances, our work suggests SNe~II cosmology can be complementary, enabling even more precise measurements of the cosmological parameters.
 
\begin{figure}
\centering
\includegraphics[width=9.5cm]{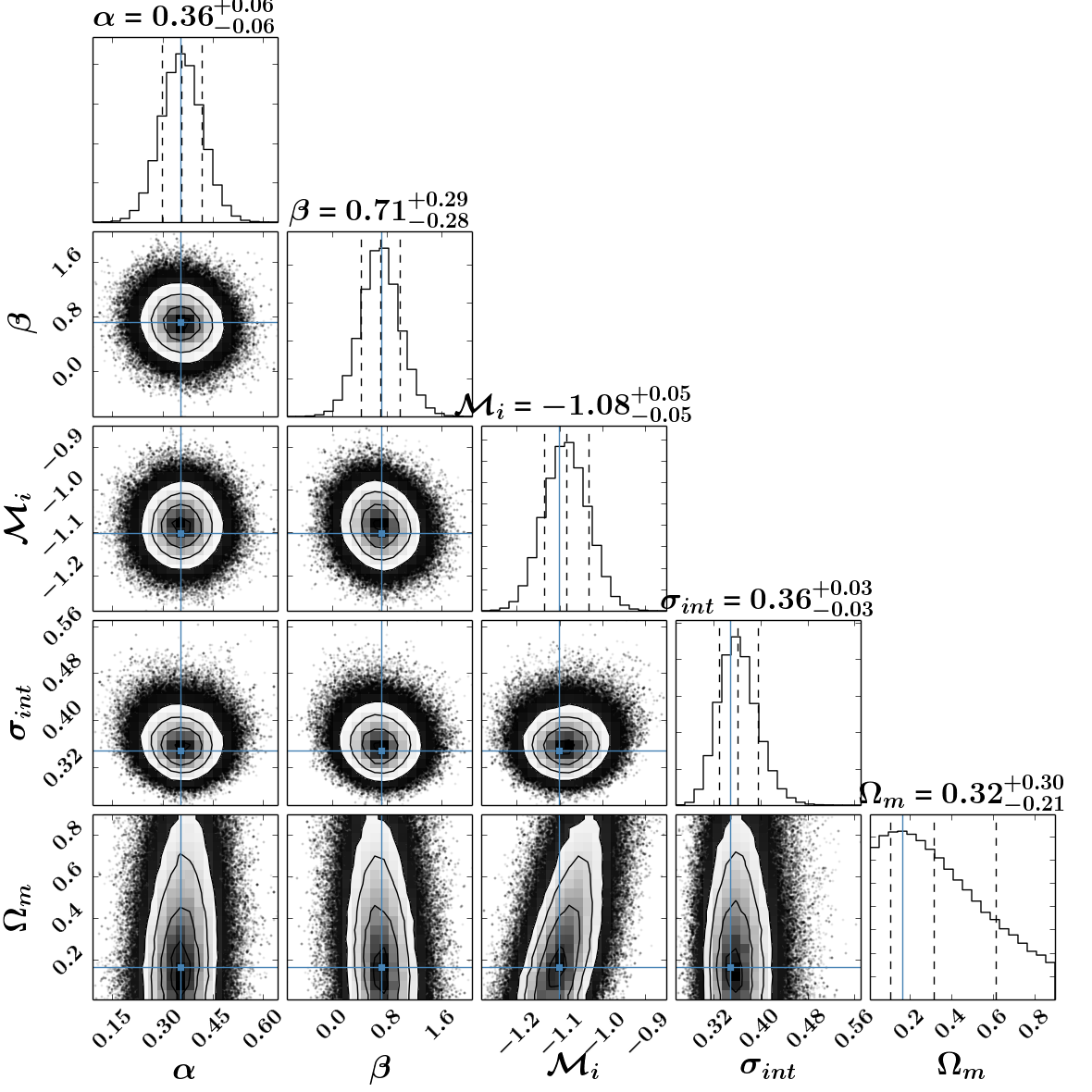}
\caption{PCM: Corner plot showing all the one and two dimensional projections. The blue lines are the values obtained using only one likelihood minimisation. Contours are shown at 1, 2, and 3 sigmas. The five free parameters are plotted: $\alpha$, $\beta$, $\mathcal{M}_{\lambda 1}$, $\sigma_{int}$, and $\Omega_{m}$. To make this figure we use the corner plot package (triangle.py v0.1.1. Zenodo. 10.5281/zenodo.11020)} 
\label{HD_PCM_cosmo}
\end{figure}

\section{Standard Candle Method}\label{SCM}

\subsection{Photospheric expansion velocities}\label{velocity_Hb}

The SCM is the most used to standardise SNe~II. This method is based on the correlation between the photospheric expansion velocities ($v_{phot}$) and the intrinsic luminosity and so requires at least one spectrum, unlike the PCM. The precise measurement of the $v_{phot}$ is not possible because no spectral line is directly connected to this velocity. However, an estimation of $v_{phot}$ (5-10\% of accuracy, \citealt{dessart05}) can be obtained through the minimum flux of the absorption component of P-Cygni line profile of an optically thin line formed by pure scattering such as \ion{Fe}{2} $\lambda$5018 or \ion{Fe}{2} $\lambda$5169.\\

\indent Measuring the \ion{Fe}{2} absorption line for noisy or early ($\leq$20 days) spectra can be very difficult, and therefore some authors attempt to use stronger features. \citet{nugent06} proposed to use the H$\beta$ $\lambda$4861 absorption line which is stronger than the weaker \ion{Fe}{2} absorption line but also present in the early spectrum. The original correlation between the H$\beta$ $\lambda$4861 and the \ion{Fe}{2} velocities found by \citet{nugent06} was revisited recently by \citet{poznanski10} and \citet{takats12}. Using 28 spectra ranging between 5 of 40 days, \citet{poznanski10} found v$_{Fe_{II}}$= 0.84 $\pm$0.05 v$_{H\beta}$, a relation confirmed by \citet{takats12} who found using the same range (between 5 of 40 days after the explosion) v$_{Fe_{II}}$=0.823 $\pm$0.015 v$_{H\beta}$. Using our spectral library at low redshift (CSP-I sample), and $\sim$ 100 spectra between 0 and 40 days after the explosion, we derive a very consistent relation. As we can see in Figure \ref{V_Hbeta_V_Fe} where we represent the \ion{Fe}{2} $\lambda$5018 velocity versus H$\beta$ $\lambda$4861 velocity, we obtain a strong correlation with a Pearson factor of 0.92 and a relation between both velocities defined as v$_{Fe_{II}}$= 0.83 $\pm$ 0.04 v$_{H\beta}$ consistent with previous studies. Note that in this work, we use \ion{Fe}{2} $\lambda 5018$ line instead of \ion{Fe}{2} $\lambda 5169$ because the latter can be blended by other elements such as the Fe triplet or \ion{Sc}{1}\\

\begin{figure}
\centering
\includegraphics[width=9.0cm]{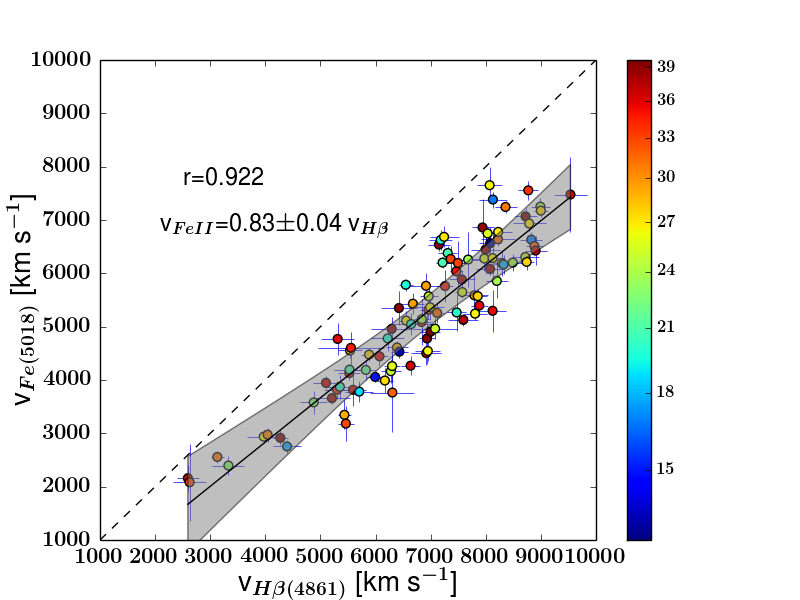}
\caption{We plot the velocities determined from the absorption minima of \ion{Fe}{2} $\lambda$5018 and H$\beta$ $\lambda$4861. The dashed line represents x=y. In this figure, only the spectra of SNe~II from the CSP-I sample at phases of 0-40 days after the explosion are plotted, i.e., 98 spectra. The shaded area is the 1$\sigma$ confidence interval using Scheffe's method. The colour bar on the right side represents the different epochs from 0 to 40 days after the explosion.}
\label{V_Hbeta_V_Fe}
\end{figure}

\indent We use H$\beta$ $\lambda$4861 velocities to standardise the SNe~II because the majority of the high redshift spectra (SDSS-II and SNLS samples) are noisy and taken at early phases where the \ion{Fe}{2} absorption lines are not visible. Errors on H$\beta$ velocities were obtained by measuring many times the minimum of the absorption changing the continuum fit. Both quantities are listed in Table \ref{SN_sample}. To find the best epoch to use the SCM we need the velocities for different epochs. As proposed by \citet{hamuyphd} and used in all the SNe~II cosmology works \citep{nugent06,poznanski09,andrea10,poznanski10,olivares10,rodriguez14,dejaeger15b} we do an interpolation/extrapolation using a power law of the form:
\begin{equation}
V(t) = A \times t^{\gamma},
\label{velocity}
\end{equation}
where $A$ and $\gamma$ are two free parameters obtained by least-squares minimisation for each individual SN and $t$ the epoch since the explosion. To derive the velocity error following the work done by \citet{dejaeger15b}, a Monte Carlo simulation is performed, varying randomly each velocity measurement according to the observed velocity uncertainties over more than 2000 simulations. Following \citet{poznanski09}, we add to the velocity uncertainty of every SN~II a value of 150 km s$^{-1}$, in quadrature, to account for unknown host-galaxy peculiar velocities. For the SNe~II with one spectrum the same power law is used but this time with a fixed $\gamma$, that is derived using only the CSP-I sample for which we have many spectra per SN and a better fit can be achieved. We find a median value of $\gamma$ $=$-0.407$\pm$ 0.173. It is important to note that in the majority of other SN~II cosmology works, the authors used the same power law for all the SNe, whereas in our work the $\gamma$ is different for all SNe~II with more than two spectra. Additionally, in Section \ref{alpha_gamma}, we show the possibility of using a new relation between $A$ and $\gamma$ in order to derive the velocity when only one spectrum is acquired without assuming the same power-law exponent.

\subsection{Methodology}

To plot the Hubble diagram, as in Section \ref{PCM}, we run a MCMC calculation and minimise the negative log of the same likelihood function (equation \ref{likelihood}) but now using another model where instead of s$_{2}$ we have now H$\beta$ velocities: 

\begin{equation}
\begin{split}
m^{model}_{\lambda 1}=&\mathcal{M}_{\lambda 1}-\alpha  log_{10}\left(\frac{v_{H\beta}}{<v_{H\beta}>~km~s^{-1}}\right) \\ +& \beta_{\lambda 1}  (m_{\lambda 2}-m_{\lambda 3}) + 5  log_{10} (\mathcal{D}_{L}(z_{CMB}|\Omega_{m},\Omega_{\Lambda})),
\end{split}
\label{scm_equa}
\end{equation}

\noindent where $\mathcal{D}_{L}(z|\Omega_{m},\Omega_{\Lambda})$, $z_{CMB}$, $\mathcal{M}_{\lambda 1}$, $\alpha$, and $\beta_{\lambda_{1}}$ are defined in the previous section and as $\sigma^{2}_{tot}$ is defined as:

\begin{equation}
\begin{split}
\sigma^{2}_{tot}=&\sigma^{2}_{m_{\lambda 1}} + (\frac{\alpha}{ln 10}\frac{\sigma_{v_{H\beta}}}{v_{H\beta}})^2 + (\beta \sigma_{(m_{\lambda 2}-m_{\lambda 3})})^2\\&+\left (\sigma_{z} \frac{5(1+z)}{z(1+z/2)ln(10)}\right )^{2}+\sigma^{2}_{int} .
\end{split}
\end{equation}

Note that equation \ref{scm_equa} is the same used by \citet{andrea10} and \citet{poznanski09} but they used the expansion velocity measured from the \ion{Fe}{2} line instead of using the H$\beta$ line as we do. As for the PCM, we center the velocity and colour distributions, i.e, we divide the distribution by the mean velocity ($< v_{H\beta}>$ $\sim$ 5900 km s$^{-1}$) and mean colour ($<(m_{\lambda 2}-m_{\lambda 3})>$ $\sim$ -0.02) of the whole sample respectively.

\subsection{Results}

\subsubsection{Fixed cosmology}

The same colour term as the PCM is used, and we plot the Hubble diagram for an epoch of 45 days in the rest-frame post-explosion. This epoch is the one with the smallest $\sigma_{int}$ and is consistent with 50 days in the rest-frame post-explosion used by other SN~II cosmology works. Our sample at this specific epoch and combination is composed of 61 SNe~II. We find an intrinsic dispersion of 0.27 mag, i.e., 12\% in distance errors. The use of the SCM allows us to reduce the intrinsic scatter from 0.55 mag (raw magnitudes) to 0.27 mag, i.e., an improvement of 13\% in distance errors. We derive $\alpha = 3.18^{+0.41}_{-0.40}$, $\beta = 0.97^{+0.26}_{-0.25} $, and $\mathcal{M}_{\lambda 1}=-1.13^{+0.04}_{-0.04}$. Assuming a Hubble constant of $H_{0}$=70 km s$^{-1}$ Mpc$^{-1}$, an absolute magnitude $M_{i}= -16.91^{+0.04}_{-0.04} $ is obtained from $\mathcal{M}_{\lambda 1}$. The Hubble diagram and its associated Hubble residual are plotted in Figure \ref{HD_SCM}. As we can see using this method we are only able to reach redshift around 0.2 where the distinction between cosmological models is very small.\\

\indent We performed the same residual analysis between the data and the $\Lambda$CDM cosmological model and find the same conclusions: no autocorrelation, no outliers according to the Chauvenet's criterion and finally, we cannot reject the null hypothesis that the residuals come from a Gaussian distribution.

\begin{figure}
\includegraphics[width=9.0cm]{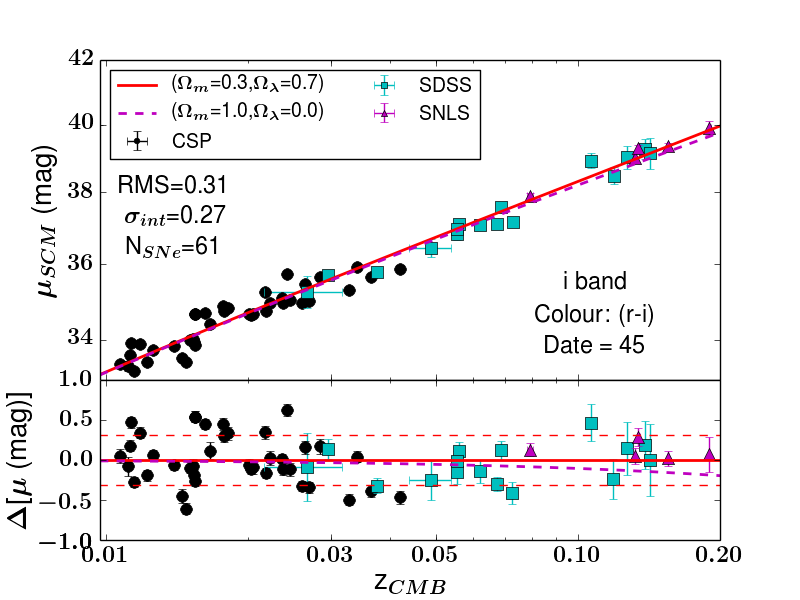}
\caption{Hubble diagram for SNe~II, using the SCM and all the SNe~II available at this epoch from the CSP-I, SDSS-II, and SNLS sample. Black dots represent the SNe~II from the CSP-I whereas the cyan squares and magenta triangles are the SDSS-II and the SNLS sample respectively. Red line is the Hubble diagram for the $\Lambda$CMB ($\Omega_{m}$=0.3 and $\Omega_{\Lambda}$=0.7) and in magenta line for an Einstein-de Sitter cosmological model ($\Omega_{m}$=1.0 and $\Omega_{\Lambda}$=0.0). In both models, we assume a Hubble constant of 70 km s$^{-1}$ Mpc$^{-1}$. In the bottom panel, the residuals with respect to the $\Lambda$CMB are shown. We also present the number of SNe~II available at this epoch (N$_{SNe}$), the epoch after the explosion (Date), the Root Mean Square (RMS) and the intrinsic dispersion ($\sigma_{int}$).}
\label{HD_SCM}
\end{figure}

\subsubsection{$\Omega_{m}$ derivation}\label{scm_cosmo}

As done in Section \ref{omega_m_PCM}, in this section, we try to derive cosmological parameters. For the SCM the highest redshift used is too small to put constraints on the dark energy density and matter density. Despite this, the same MCMC calculation done in the Section \ref{omega_m_PCM} for a flat Universe is performed. Figure \ref{HD_SCM_cosmo} presents the same as Figure \ref{HD_PCM_cosmo} but this time using the SCM. We find values consistent with that found with a fixed cosmology: $\alpha = 3.18^{+0.41}_{-0.41}$, $\beta = 0.97^{+0.26}_{-0.25} $, and $\mathcal{M}_{\lambda 1}=-1.13^{+0.04}_{-0.04}$, and $\sigma_{int}=0.29^{+0.03}_{-0.03}$. \\
\indent For the matter density we see a less pronounced drop than that obtained using the PCM. The value derived for the matter density is $\Omega_{m}= 0.41^{+0.31}_{-0.27}$, which corresponds to $\Omega_{\lambda}=0.59^{+0.27}_{-0.31}$. In Figure \ref{HD_SCM_cosmo} the blue lines represent the value derived using a simple likelihood minimisation (without MCMC), e.g. for the density matter we obtain $\Omega_{m}$= 0.20 $\pm$ 0.49. The difference in the matter density error between the SCM and the PCM (0.49 versus 0.30) is not due to the method (the intrinsic dispersion is better for the SCM) but is due to the redshift range and the number of SNe~II. We clearly require higher redshift SNe~II (z$\geq$0.3) to derive cosmological parameters and obtain better constraints on the matter density.

\begin{figure}
\includegraphics[width=9.5cm]{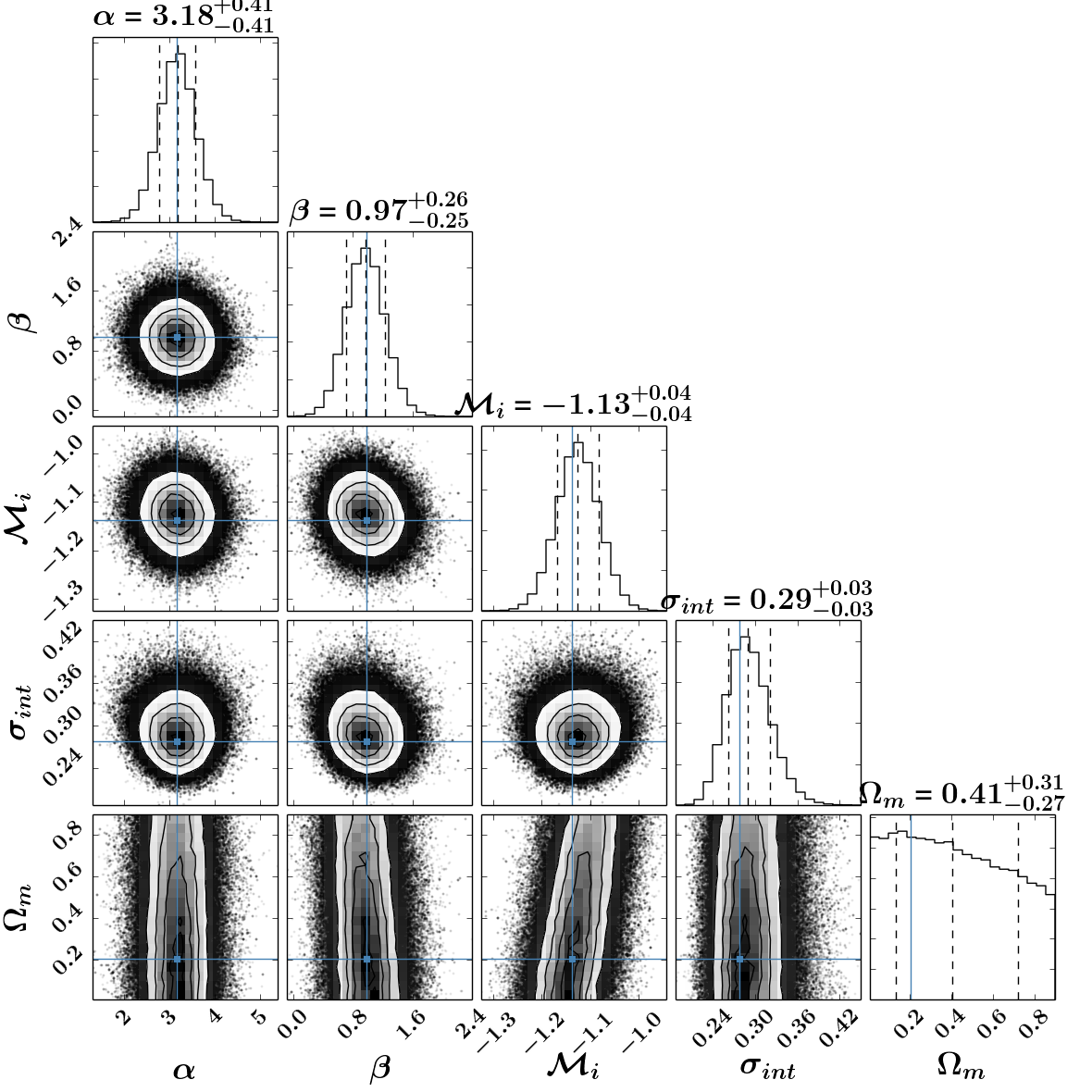}
\centering
\caption{SCM: Corner plot showing all the one and two dimensional projections. The blue lines are the values obtained using only one likelihood minimisation. Contours are shown at 1, 2, and 3 sigmas. The five free parameters are plotted: $\alpha$, $\beta$, $\mathcal{M}_{\lambda 1}$, $\sigma_{int}$, and $\Omega_{m}$}
\label{HD_SCM_cosmo}
\end{figure}
\twocolumngrid

\section{Discussion}

\subsection{Sample Comparison} \label{com_sample}

In this part, we will compare the three samples used for this work and with the SCM: CSP-I, SDSS-II, and SNLS. In Figure \ref{distri} (top), we compare the absolute magnitude uncorrected for velocity or host extinction and assuming a standard cosmology ($\Omega_{m}=0.3$, and $\Omega_{\Lambda}=0.7$). Even if the number of SNe~II used is very different (40 for CSP-I, 16 for SDSS-II, and 5 for SNLS), the luminosity distribution appears different. The CSP-I sample has absolute magnitudes over a range of 2 magnitudes which is expected for SNe~II. For the SDSS-II sample, as found by \citet{andrea10}, it spreads only a small range of absolute magnitude (0.7 mag). The authors explained the lack of any dim SNe~II above $z$=0.10 by the Malmquist bias. At low redshift, we do not have intrinsically dimmer SNe~II in the SDSS-II sample due to the fact that dimmer candidates were not followed spectroscopically. For the SNLS sample, the statistic is too low to derive conclusions. The same result is found by analysing the distribution of H$\beta$ velocities. The CSP-I sample spreads a large range of velocities (2500-8500 km s$^{-1}$) while the SDSS-II sample has in general high velocities (5000-8000 km s$^{-1}$). The lack of low velocities for the SDSS-II sample could be explained by the bias toward more luminous SNe~II in the SDSS-II sample. This bias could explain the higher dispersion in the Hubble diagram for the low-redshift SNe~II. SNe~II from the CSP-I sample spread a larger range in observed properties than the SDSS-II sample which are biased toward more luminous events. In the future, with larger datasets and simulations (see Section 6.7 for the Malmquist bias), we will be able to better characterise systematic biases.

\begin{figure}
\includegraphics[width=9.0cm]{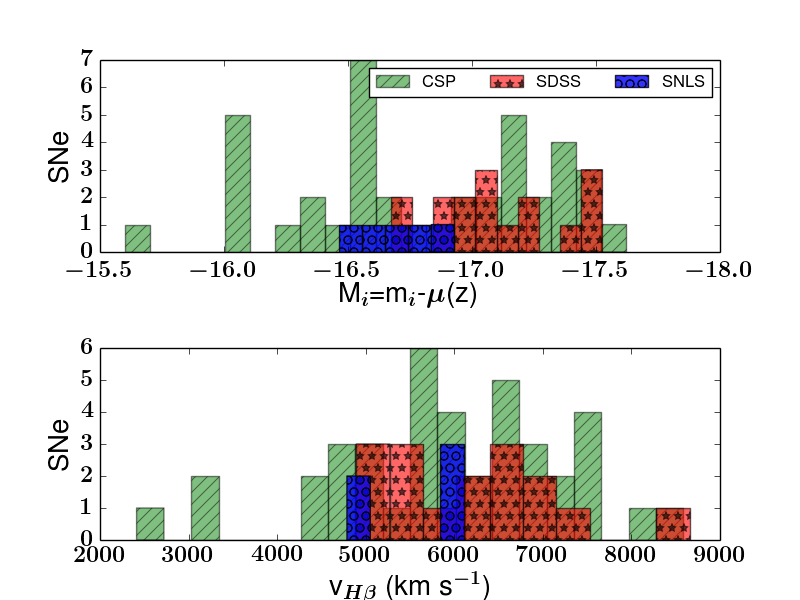}
\caption{Comparison of the CSP-I, SDSS-II and SNLS samples. Top panel represents the absolute magnitude without any calibration (not corrected for velocity or dust) and assuming a Hubble constant of 70 km s$^{-1}$ Mpc$^{-1}$, $\Omega_{m}=0.3$, and $\Omega_{\Lambda}=0.7$. The bottom panel shows the distribution of H$\beta$ velocity. In both plot, the green colour represents the CSP-I sample, the red the SDSS-II sample, and the blue the SNLS sample.}
\label{distri}
\end{figure}

\subsection{PCM versus SCM}

Using a larger data sample and higher redshift SNe~II than \citealt{dejaeger15b} ($\sim$ 40 SNe~II up to $z$$\sim$0.04 with $\sigma_{int}$ = 0.41 mag), we obtain an intrinsic dispersion of 0.35 mag with the PCM and 0.27 mag with the SCM. The SCM is a better method to standardise the SNe~II in term of intrinsic dispersion, but the difference between both methods is only of 0.08 mag, i.e., 3\% in distances. In contrary to the SCM, with the PCM, we are able to use more SNe~II (73 versus 61) and it can be extended to higher redshifts ($\sim$ 0.5 versus $\sim$ 0.2). The next generation of telescopes will observe many thousands of SNe~II and the PCM will be very useful to derive cosmological parameters. In Figure \ref{SCM_PCM} we present the distance modulus obtained using the PCM and the SCM. For these two methods, we have 59 SNe~II in common. As we can see the values derived are very consistent using both methods with a RMS of 0.29 mag. All the distance moduli calculated for both methods are listed in Table~\ref{SN_sample}.

\begin{figure}
\includegraphics[width=9.0cm]{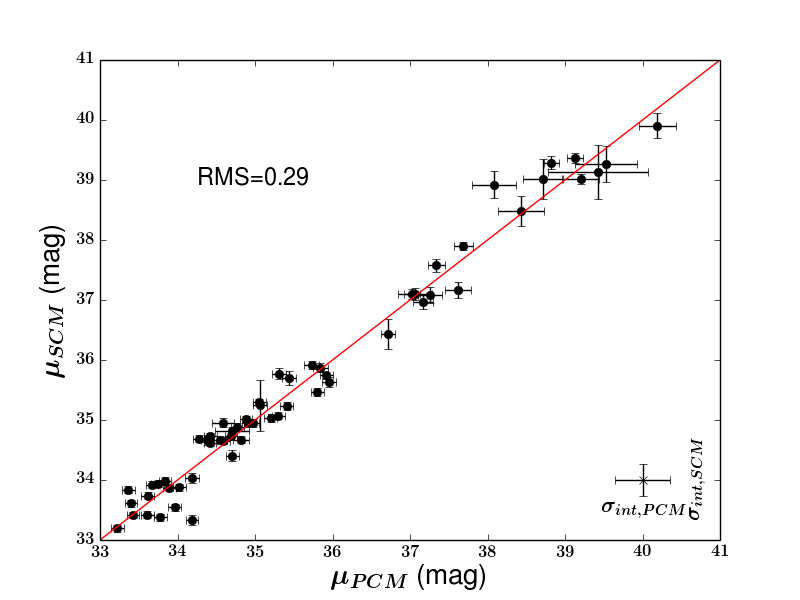}
\caption{Comparison distance modulus obtained using the PCM (x-axis) and the SCM (y-axis). The red line shows $x=y$. Note that the error bars do not include the intrinsic scatter ($\sigma_{int}$) of each method which are represented by the cross on the bottom right of the figure.}
\label{SCM_PCM}
\end{figure}

\subsection{SCM versus others works}

\indent The scatter found in this work is very consistent with those found by previous studies \citep{nugent06,poznanski09,andrea10,poznanski10,olivares10,rodriguez14,dejaeger15b}. For example with a similar methodology (same likelihood), \citet{andrea10} using 15 SNe~II from SDSS-II with 34 low-redshift SNe~II from \citet{poznanski09}, found an intrinsic dispersion of 0.29 mag. They also derived consistent free parameters $\alpha = 4.0 \pm 0.7 $, $\beta = 0.8 \pm 0.3$ but a different absolute magnitude $M_{I}=-17.52 \pm 0.08$ mag. This is largely because they assumed an intrinsic colour of 0.53 mag to correct their magnitudes for extinction.\\
\indent \citet{poznanski09} also found similar dispersion, i.e., 0.38 mag using 40 SNe~II (``full'') or 0.22 mag after removing six outliers (``culled''). In Table \ref{SCM_param} we present the values of $\alpha$, $\beta$, M$_{I}$, and $\sigma_{int}$ from different works (values taken in Table 4 of \citealt{andrea10}) and using our different samples (CSP-I, SDSS-II, SNLS). As we can see from this table, even if the free parameters are consistent with our work, small differences are present. For example, the discrepancy in the value of $\alpha$ could be explained by the method used. In this paper, we use the H$\beta$ velocity while both other studies used the iron line. We also calculate a power-law for the majority of the SNe~II for the velocity while both authors assumed a unique power-law for all SN. Thus, these differences could have an impact on the $\alpha$ value. For the $\mathcal{M}_{\lambda 1}$ as stated previously we do not correct the colour for intrinsic colour which could affect the value derived for the $\mathcal{M}_{\lambda 1}$. Additionally, the discrepancies on $\beta$ and the $\mathcal{M}_{\lambda 1}$ could arise from differences in the filters used. They used the Bessel filters $R$ and $I$ while we used the CSP-I filters $r$, and $i$.\\
\indent Note also, that the differences of methodology are not the only cause affecting the SCM fit parameters. \citet{andrea10} explained that these effects could arise from selection effects as described in Section \ref{com_sample}\\

\indent We can also compare our distance moduli with those derived by \citet{poznanski09}. \citet{poznanski09} sample and our share two SNe~II: 04D1pj and 04D4fu. \citet{poznanski09} derived a distance modulus of 39.28 $\pm$ 0.11 and 38.85 $\pm$ 0.11 while we obtain 39.367 $\pm$ 0.084 and 39.018 $\pm$ 0.087 for 04D1pj and 04D4fu respectively. These two values are very consistent.

\begin{table*}[ht]
\caption{PCM-fit Parameters.}
\centering
\begin{tabular}{cccccc}
\hline
\hline
Data Set & $\alpha$  & $\beta$ &$M_{i}$ &$\sigma_{int}$ &SNe\\
\hline
\citet{poznanski09} ``full'' &4.4 $\pm$ 0.7 &0.6$^{+0.3}_{-0.4}$ &$-$17.42 $\pm$0.10 &0.35 &40\\
\citet{poznanski09} ``culled'' &4.2 $\pm$ 0.6 &0.8$^{+0.3}_{-0.3}$ &$-$17.38 $\pm$0.08 &0.20 &34\\
SDSS-II from \citet{andrea10} &1.8$^{+0.9}_{-1.0}$ &0.1 $\pm$ 0.5 &$-$17.67$^{+0.11}_{-1.0}$ &0.16 &15\\
\citet{andrea10} + \citet{poznanski09}``culled'' &4.0 $\pm$ 0.7 &0.8$^{+0.3}_{-0.3}$ &$-$17.52 $\pm$0.08 &0.29 &49\\
CSP-I &3.04$^{+0.48}_{-0.47}$ &1.54$^{+0.38}_{-0.37}$ &$-$16.85$^{+0.05}_{-0.05}$ &0.31$^{+0.04}_{-0.03}$ &40\\
CSP-I$+$SDSS-II &3.16$^{+0.42}_{-0.42}$ &1.01$^{+0.28}_{-0.27}$ &$-$16.92$^{+0.05}_{-0.05}$ &0.30$^{+0.03}_{-0.03}$ &56\\
CSP-I$+$SNLS &3.05$^{+0.45}_{-0.44}$ &1.41$^{+0.35}_{-0.35}$ &$-$16.84$^{+0.05}_{-0.05}$ &0.30$^{+0.04}_{-0.03}$ &45\\
SDSS-II$+$SNLS &3.55$^{+0.82}_{-0.77}$ &0.39$^{+0.25}_{-0.24}$ &$-$17.02$^{+0.05}_{-0.05}$ &0.18$^{+0.05}_{-0.04}$ &21\\
CSP-I$+$SDSS-II$+$SNLS &3.18$^{+0.41}_{-0.41}$ &0.97$^{+0.26}_{-0.25}$&$-$16.91$^{+0.04}_{-0.04}$ & 0.29$^{+0.03}_{-0.03}$ &61\\
\hline
\multicolumn{5}{l}{%
\begin{minipage}{12.5cm}%
Note: Best-fit values and the associated errors for each parameter for different samples using the SCM.%
\end{minipage}%
}\\
\end{tabular}
\label{SCM_param}
\end{table*}

\subsection{H$\beta$ velocity: A and $\gamma$ correlation}\label{alpha_gamma}

\citet{Pejcha15b} found a correlation between the two free parameters (A and $\gamma$) used in the expansion velocity formula described in equation \ref{velocity}. They found that velocity decays faster in SNe~II with initially higher velocity. Using all the SNe~II from the CSP-I sample with more than three spectra (46 SNe~II), we present in Figure \ref{A_gamma} the plot of the power-law exponent ($\gamma$) versus the initial velocity (A). As we can see, our observational data confirm the result found by \citet{Pejcha15b}: SNe~II with high initial velocity decay faster. Additionally, we remark that the shape of both relations (from \citealt{Pejcha15b} and ours) is very consistent. They found a bi-modal correlation, but with $\gamma$ lower because in their model a constant velocity offset is added. This typically makes $\gamma$ more negative. Note that we find similar correlation factor, -0.82 and - 0.86 for their work and our study respectively. The relation between these quantities is very important to derive the expansion velocity for the SNe~II with only one spectrum. In the literature, the majority of the studies assumed the same power-law exponent for all SNe~II or assumed a median value for the SNe~II with only one spectrum (as done in Section \ref{velocity_Hb}). However thanks to this relation, we can derive the $H_{\beta}$ velocity with more accuracy. In Figure \ref{A_gamma} we show four different fits: a power-law (black), a linear fit (blue), an inverse fit (green), and a bi-modal fit (cyan). The best reduced chi square and dispersion are obtained using the bi-modal fit (16 and 0.08 respectively). If the $H_{\beta}$ velocities for the SNe~II with only one spectrum are derived using the two lines fit, i.e., $\gamma$=$-$1.71 $\times$ 10$^{-5}$ A $+$ 5.25 $\times$ 10$^{-2}$ for A$\leq$30500 or $\gamma$=$-$3.82 $\times$ 10$^{-6}$ A $-$0.35 for A$\geq$30500, we are able to derive a Hubble diagram with an equivalent dispersion ($\sigma_{int}$ $\sim$ 0.28 mag) to that derived in Section \ref{scm_cosmo}.

\begin{figure}
\includegraphics[width=9.0cm]{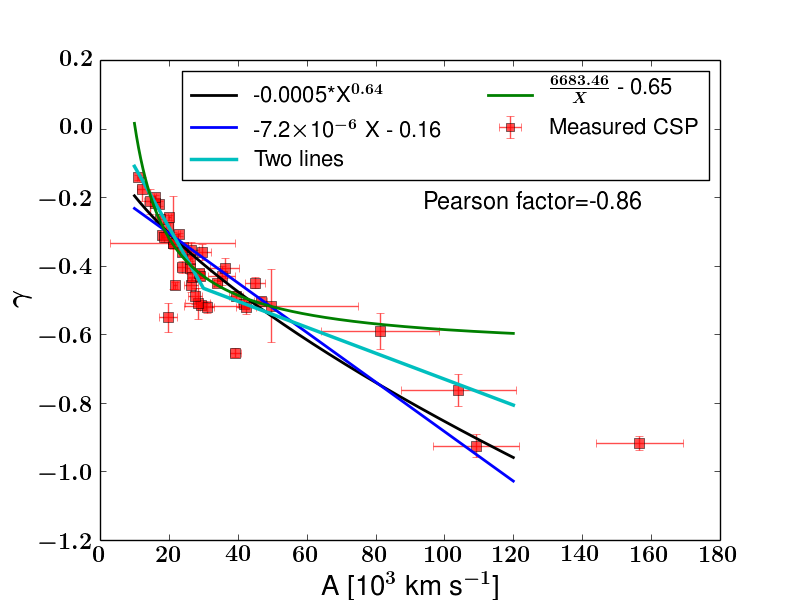}
\caption{$\gamma$ versus A using the CSP-I sample. The red squares represent the CSP-I sample, the black line a power-law fit, the blue line is a linear fit, the green line is an inverse fit, and in cyan a two lines fit.}
\label{A_gamma}
\end{figure}

\subsection{Sensitivity to progenitor metallicity?}

Using theoretical models, \citet{kasen09} suggested that progenitors with different metallicities ($Z=$0.1-1 Z$_{\odot}$) could introduce some systematic errors (0.1 mag) in the photospheric expansion velocity-intrinsic luminosity relation. From an observational point of view, \citet{anderson16a} and \citet{taddia16a} using SNe~II from the CSP-I and the intermediate Palomar Transient Factory (iPTF), respectively, showed that the equivalent width of the \ion{Fe}{2} (EW$_{Fe}$) and the absolute magnitude at maximum peak are correlated in the sense that SNe~II with smaller EW$_{Fe}$ tend to be brighter.\\ 
\indent In this part, we aim to study this using only the CSP-I sample, which is the only available sample where metal line measurements are possible. We linearly interpolate the equivalent width to 45 days post-explosion and for this specific epoch, and end up with a sample of 25 SNe~II. Note that a MC simulation is performed varying randomly each EW$_{Fe}$ measurement according to their uncertainties and linearly interpolate at epoch 45 days post-explosion and then we take as the final EW$_{Fe}$ the median while the error is the standard deviation of these 2000 fits.\\ 
\indent Figure \ref{metallicity} shows EW$_{Fe}$ versus the absolute Hubble diagram residual to the $\Lambda$CDM model ($\Omega_{m}$=0.3 and $\Omega_{\Lambda}$=0.7) and using the SCM. We find a trend between the EW$_{Fe}$ and the absolute residual, i.e., SNe~II with smaller EW$_{Fe}$ have less dispersion. The Pearson factor of 0.41 confirmed this tiny relation. This figure could reflect the existence of one category of SNe~II more standardisable than other, i.e., SNe~II with small EW$_{Fe}$ ($<$ 10 $\AA$) seem to be better standard candles than the others. It will be very interesting to construct a Hubble diagram using only SNe~II with small EW$_{Fe}$, but unfortunately a sufficient number of SNe~II are unavailable. If the Hubble diagram residual is taken instead of the absolute of the Hubble diagram residual no correlation is found with the EW$_{Fe}$. Note that in our Hubble diagram (Figure \ref{HD_SCM}), the higher redshift SNe~II (SDSS-II, SNLS) seem to have less intrinsic dispersion than the low-redshift sample. This could be also explained by the fact that higher redshift SNe~II have a smaller range in luminosity (Section \ref{com_sample}), thus, a smaller range in EW$_{Fe}$, which could imply less scatter.\\

\indent If we use the equivalent width of the \ion{Fe}{2} $\lambda 5018$ absorption line as proxy for metallicity \citep{dessart14}, and if the Hubble diagram residual is only coming from the metallicity, we can conclude as \citet{kasen09} that differences in metallicity introduce some scatter in the Hubble diagram. All the figures and discussions regarding metallicity will be left for a future publication (Gut\'ierrez et al. in preparation).

\begin{figure}
\includegraphics[width=9.0cm]{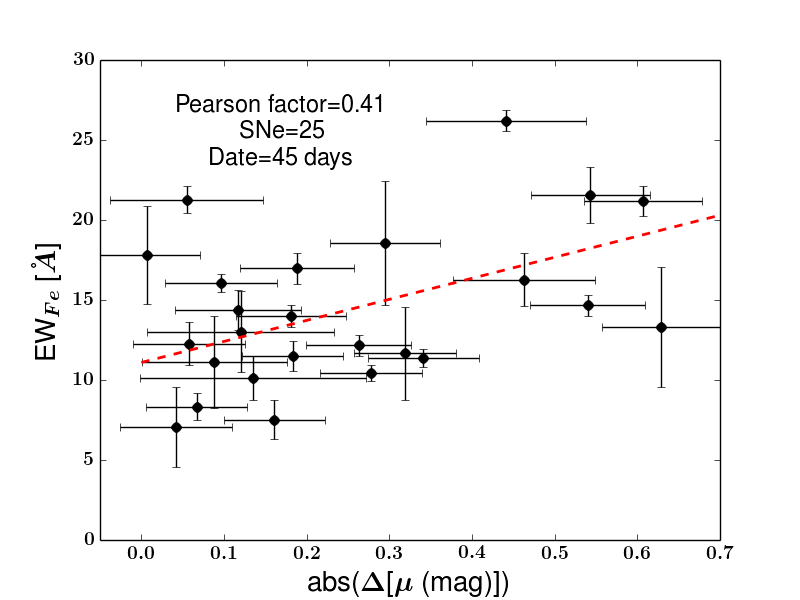}
\caption{The absolute Hubble residual using the SCM and the $\Lambda$CDM cosmological model versus the equivalent width of the \ion{Fe}{2} $\lambda 5018$ absorption line. The red dashed line represents the linear regression taking into account errors in both variables using the Python port of B. Kelly's LINMIX ERR IDL package (\citealt{kelly07}, https://github.com/jmeyers314/linmix.}
\label{metallicity}
\end{figure}

\subsection{Combined SNe~Ia and SNe~II Hubble diagram}

In this section, we combine our SNe~II sample with a complete SNe~Ia sample (740 SNe~Ia) from \citet{betoule14}. In Figure \ref{Ia_II}, we show the combined Hubble diagram where both samples are fitted separately, i.e., using the distance modulus derived with the PCM for the SNe~II and the distance modulus using the fitting parameters from \citet{betoule14} for the SNe~Ia sample. Then, we try to fit simultaneously both samples following the work done by \citet{Scovacricchi2016} where they combined simulated 100 SuperLuminous SNe from SUDSS and 3800 SNe~Ia from DES. We minimise the likelihood corresponding to the product of two likelihoods $\mathcal{L}=\mathcal{L}_{Ia}*\mathcal{L}_{II}$. We have thus 9 free parameters: $\alpha$, $\beta$, $\mathcal{M}_{\lambda 1}$, and $\sigma_{int}$ for the two likelihoods plus the same $\Omega_{m}$. Note that for the SNe~Ia sample, we use the same likelihood used in equation \ref{likelihood} but, instead of the $s_{2}$ value or the H$\beta$ velocity we use the stretch parameter. To estimate the effect of combining the two samples, we look at the value derived for $\Omega_{m}$ and especially its uncertainty. Using the PCM or the SCM the precision derived for the matter density with the combined samples is not better than the one obtained using only the SNe~Ia sample. This can be easily explained by three different factors: the redshift range (SNe~Ia up to 1.2), the size of the sample, and the fact that SNe~Ia are better standardisable ($\sigma_{int}$ $\sim$ 0.10-0.15 mag).\\
\indent To compare the difference in precision achieved with the SNe~Ia and SNe~II, we restrict the SNe~Ia sample to the same SNe~II redshift range, i.e., $z\leq0.5$. Doing a Monte Carlo simulation (hundred iterations), 73 SNe~Ia (equivalent to the SNe~II sample size) are randomly selected. A median uncertainty in the matter density of 0.1 is derived which compares to the $\sim$ 0.3 using only SNe~II. Otherwise, we can count how many SNe~Ia are necessary to reach a precision of $\sim$ 0.3 in the density matter comparing to the 73 SNe~II needed. We find that 22 SNe~Ia or 13 SNe~Ia are required using the PCM and the SCM respectively which corresponds to $\sim$ 30 \% ($\pm$ 8\%) or $\sim$ 20 \% ($\pm$ 7\%) of the SNe~II sample size for the PCM and the SCM respectively.
Individually, even if the SNe~Ia are better standard candles than the SNe~II, SNe~II cosmology can provide an independent measurement of the cosmological parameters. Or, with growing samples in the future, they may be used, as shown here, in a combination with SNe~Ia. As stated earlier, SN~II progenitors are better understood than those of SNe~Ia (given their direction detection on pre-explosion images, e.g. \citealt{smartt09a}), which may allow us to further reduce the intrinsic dispersion, possibly reaching the same dispersion offered by SNe~Ia.

\begin{figure}
\includegraphics[width=9.0cm]{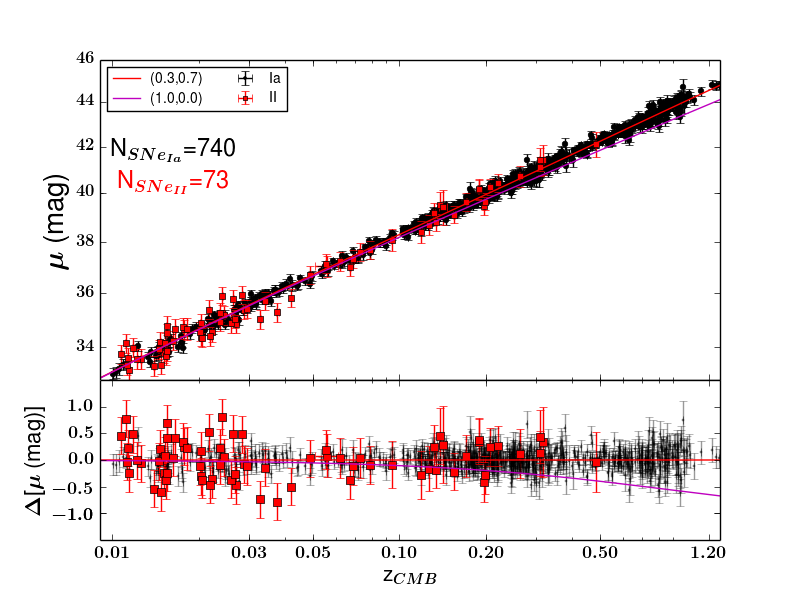}
\caption{Combined Hubble diagram using SNe~Ia from \citet{betoule14} and SNe~II. The red squares are for the SNe~II (using the PCM) and the black dots for the SNe~Ia. Note that distance modulus errors for both methods include the intrinsic dispersion. The red line the Hubble diagram for the $\Lambda$CMB ($\Omega_{m}$=0.3 and $\Omega_{\Lambda}$=0.7) and in magenta line an Einstein-de Sitter cosmological model ($\Omega_{m}$=1.0 and $\Omega_{\Lambda}$=0.0)}
\label{Ia_II}
\end{figure}

\subsection{Malmquist bias}

Using a basic simulation, we investigate the Malmquist bias, which leads to observe preferentially brighter objects in a magnitude limited survey. This bias implies a mean observed magnitude brighter than the intrinsic mean, biasing distance measurements and therefore overestimating the matter density. This definition is only valid if the magnitudes have a Gaussian distribution, which it is the case for the SNe (see \citealt{anderson14a} for magnitude distributions at different epochs.)\\
\indent To derive an approximate Malmquist bias, we calculate the difference in magnitudes after applying the PCM (s$_{2}$ and colour corrections) of fake input SNe~II distance moduli and output SNe~II distance moduli, i.e., the SNe~II which have passed the limiting magnitude cut of the survey. The ($g$, $r$, $i$, $z$) limiting magnitudes assumed are (25.5, 25.0, 24.8, 23.9) and (22.2, 22.2, 21.3, 20.5) for SNLS and SDSS-II respectively.\\
\indent To generate fake apparent magnitudes, we adopt our low-redshift sample distribution (CSP-I), i.e., the absolute $V$ band magnitudes at the end of the plateau (M$_{end}$) between $-$15.0 and $-$17.0 \citep{anderson14a}. Using MC simulations and the correlation matrix between M$_{end}$, the s$_{2}$, the plateau duration (Pd), and the colours $(B-V)$, we generate 10 000 mock SNe~II that follow the nearby distribution. {Due to the fact that it is a simplistic simulation, the effects of observing conditions (such as seeing, sky transparency, etc) are not modelled. However, we do take into account the same magnitude errors of the actual observed SNe of the CSP-I survey and inject them into the fake SNe. Then, we select randomly 2000 SNe~II (i.e., $\sim$ 2000 set of M$_{end}$, s$_{2}$, Pd, ($B-V$)) from the MC simulation and for each SN~II we use $\sim$ 2000 random redshifts between 0.03 and 0.80. From this, we can derive an apparent magnitude (i.e. $\sim$ 4 000 000 magnitudes) assuming a $\Lambda$CDM model and applying an inverse K-correction (similar to Section \ref{procedure}). Then, we compare the apparent magnitude (at the end of the plateau) with the limiting magnitude which depends on the redshift. For a redshift $\leq$ 0.3, an absolute magnitude in $V$ band corresponds to an apparent magnitude in the $r$ band, from 0.3$<$z$<$0.56 to the $i$ band, finally from 0.56$<$z$<$0.80 to the $z$ band. All the SNe~II which pass this cut form our output sample. From the input and output SNe~II, we can derive the modulus distances after correcting the magnitude using $\alpha=0.37 \pm 0.10$ and $\beta=1.20 \pm 0.35$. The $\alpha$ and $\beta$ values were derived using only the CSP-I sample and applying the PCM as achieved in \citet{dejaeger15b}. The final Malmquist bias is taken as the mean value of the modulus distance difference of the input and output sample for a redshift bin of 0.02. Then, we interpolate linearly the Malmquist bias over all the bins in order to apply this bias to each SN of the SDSS and SNLS sample. The errors for each bin are taken as the standard error of the mean. For clarity, the standard deviation ($\sim$ 0.5 mag) is not shown in this figure.\\
\indent Roughly, for the SNLS survey, we derive a mean Malmquist bias of $\sim$ $-$0.02 mag for a redshift range between 0.02 and 0.3. After this range, the mean Malmquist bias decreases up to $\sim$ $-$ 0.16 mag for a redshift between 0.3 and 0.4 and finally to $\sim$ $-$0.32 mag for a redshift between 0.4 and 0.5. The Malmquist bias decreasing after z=0.22 can be modelled by a straight line of equation:$MB_{SNLS}=-1.21 (\pm 0.10) \times z + 0.25 (\pm 0.03)$. Similarly, for the SDSS survey, we derive a mean Malmquist bias of $\sim$ $-$0.08 mag for a redshift range between 0.02 and 0.1 and $\sim$ $-$0.31 mag for a redshift range between 0.1-0.15. As for SNLS, we fit the decreasing Malmquist bias after z=0.06 by a straight line of equation $MB_{SDSS}=-4.61 (\pm 0.55) \times z + 0.23 (\pm 0.05)$. In Figure \ref{Malmquist}, we present this approximate Malmquist bias versus redshift. The high values derived for all surveys are a warning for SNe~II cosmology. Deriving strong constraints for the cosmological parameters requires measurements extending far back in time where the Malmquist bias is important. Thus, it will be difficult to reach the same level of precision that obtained with the SNe~Ia for which the Malmquist bias is much smaller \citep{perrett10} and the intrinsic dispersion too. In the future it is crucial to obtained a good estimation of this bias with a full simulation as achieved by \citet{perrett10}.\\

\indent Even if our method is an approximation, we apply the Malmquist bias to each SN~II in our Hubble diagram. For this, to each SN~II corrected apparent magnitude, the value of the Malmquist bias is added at the SN~II redshift. Then, we derive the matter density and compare it with the value obtained in Section \ref{omega_m_PCM}. The matter density distribution shape is very similar to that derived in Figure \ref{HD_PCM_cosmo} and the value obtained, $\Omega_{m}$=0.35$^{+0.30}_{-0.22}$, is also very consistent. We caution again the reader that even if the Malmquist bias does not seems to affect our cosmology it should be calculated with more accuracy.

\begin{figure}
\includegraphics[width=9.0cm]{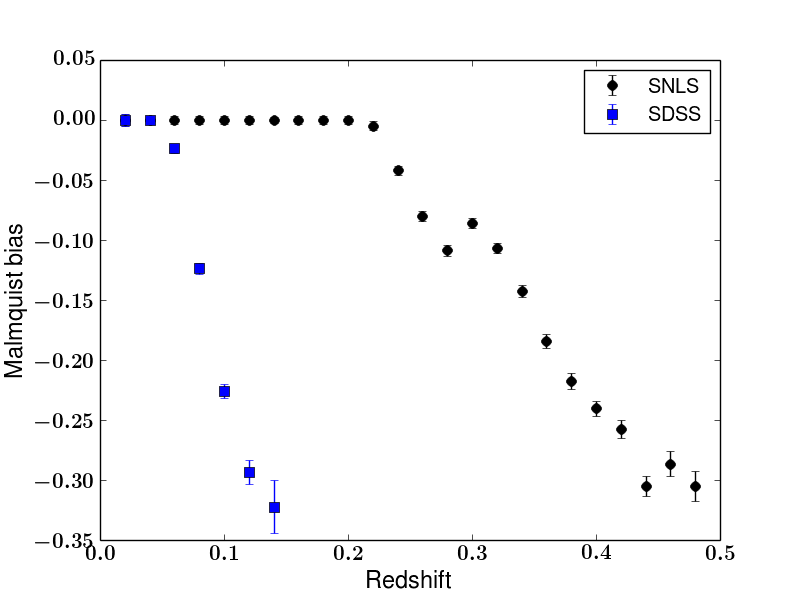}
\caption{The squares represent the simulated Malmquist bias versus redshift for the SNLS and SDSS-II survey in black and blue respectively.}
\label{Malmquist}
\end{figure}

\section{Conclusions}

Using three samples, CSP-I, SDSS-II, and SNLS, we construct the two largest SNe~II Hubble diagrams (73 SNe~II in this work versus 49 SNe~II in the literature), extending successfully the Photometric colour Method developed in \citet{dejaeger15b} to higher redshifts (up to 0.5). We also compare this method with the Standard Candle Method. In summary :
\begin{enumerate}
\item{Using the PCM we find an intrinsic dispersion of 0.35 mag (73 SNe~II with a redshift up to $\sim$ 0.5) using the $i$ band, while using the SCM we obtained a dispersion of 0.27 mag (61 SNe~II with a redshift up to $\sim$ 0.2).}
\item{The Hubble diagram derived from the CSP-I$+$SDSS-II$+$SNLS samples using the SCM yields a dispersion similar than those found in the literature and emphasises the potential of SCM in cosmology.}
\item{We find a relation between the power-law exponent ($\gamma$) and the initial velocity (A) which is very useful to derive H$_{\beta}$ velocities at any epoch with only one SNe~II spectrum.}
\item{We derived cosmological parameters ($\Omega_{m}$) consistent with the $\Lambda$CDM. Using the PCM we found $\Omega_{m}$=0.32$^{+0.30}_{-0.21}$, and with the SCM $\Omega_{m}$=0.41$^{+0.31}_{-0.27}$. These results are consistent with the existence of the dark energy at two sigma.}
\item{The distance moduli derived using the PCM and the SCM are very consistent with a dispersion of 0.29 mag.}
\item{Using a simple simulation, we point out the high values for the SNe~II Malmquist bias which could be problematic to achieve in the future, similar cosmological constraint uncertainties that those obtained with the SNe~Ia.}
\end{enumerate}
While SNe~II currently display larger scatter in their use as distance indicators as compared to SNe~Ia, in this work we have shown that SNe~II can be used as a viable independent cosmological probes. Indeed, with future large surveys predicted to significantly extend the number of SNe~II at higher redshift, these objects promise to provide a valuable sanity check to results obtained from other methods.

\acknowledgments
Support for T. D., S. G., M. H., H. K., C.G, is provided by the Ministry of Economy, Development, and Tourism's Millennium Science Initiative through grant IC120009, awarded to The Millennium Institute of Astrophysics, MAS. S.G. acknowledges support from Basal Project PFB–03. H.K. also acknowledge support by CONICYT through FONDECYT grants 3140566 and 3140563 respectively. L.G. was supported in part by the US National Science Foundation under Grant AST-1311862.M. D. S. gratefully acknowledge generous support provided by the Danish Agency for Science and Technology and Innovation realized through a Sapere Aude Level 2 grant. M.S. acknowledges support from EU/FP7-ERC grant 615929..The work of the CSP-I has been supported by the National Science Foundation under grants AST0306969, AST0607438, and AST1008343. This research has made use of the NASA/IPAC Extragalactic Database (NED) which is operated by the Jet Propulsion Laboratory, California Institute of Technology, under contract with the National Aeronautics and Space Administration and of data provided by the Central Bureau for Astronomical Telegrams. This work is based in part on data produced at the Canadian Astronomy Data Centre as part of the CFHT Legacy Survey, a collaborative project of the National Research Council of Canada and the French Centre National de la Recherche Scientifique. The Work is also based on observations obtained at the Gemini Observatory, which is operated by the Association of Universities for Research in Astronomy, Inc., under a cooperative agreement with the NSF on behalf of the Gemini partnership: the National Science Foundation (United States), the STFC (United Kingdom), the National Research Council (Canada), CONICYT (Chile), the Australian Research Council (Australia), CNPq (Brazil) and CONICET (Argentina). This research used observations from Gemini program numbers: GN-2005A-Q-11, GN-2005B-Q-7, GN-2006A-Q-7, GS-2005A-Q-11 and GS-2005B-Q-6, GS-2008B-Q-56


\end{document}